\documentclass{siamltex}
\usepackage{amsmath}
\usepackage{amssymb}
\usepackage{dsfont}
\usepackage{color}
\usepackage{graphicx}
\usepackage{subfig}
\usepackage{multirow}
\usepackage{morefloats}
\usepackage{pgfplots}
\usetikzlibrary{plotmarks}
\usepackage{tikz}

\newtheorem{remark}[theorem]{Remark}

\begin{document}
\title{Particle methods for multi-group pedestrian flow}

\author{N.K. Mahato \footnotemark[1]\and A. Klar\footnotemark[1] \footnotemark[2]
       \and S. Tiwari \footnotemark[1] }
\footnotetext[1]{Technische Universit\"at Kaiserslautern, Department of Mathematics, Erwin-Schr\"odinger-Stra{\ss}e, 67663 Kaiserslautern, Germany 
  (\{mahato,klar,tiwari\}@mathematik.uni-kl.de)}
\footnotetext[2]{Fraunhofer ITWM, Fraunhoferplatz 1, 67663 Kaiserslautern, Germany}

\maketitle
\begin{abstract}
\noindent We consider a multi-group microscopic model for pedestrian flow describing the behaviour of large groups. It is based on an interacting particle system coupled to an eikonal equation. Hydrodynamic  multi-group models are derived from the underlying particle system as well as scalar multi-group models. The eikonal equation is used to compute  optimal paths for the pedestrians. Particle methods are used to solve the equations on all levels of the hierarchy. Numerical test cases are investigated  and the   models and, in particular, the resulting evacuation times are compared for a wide range of  different parameters. 
\end{abstract}


\section{Introduction}
Pedestrian flow modelling  has attracted the interest of a large number of scientists from different research fields, as well as  planners and designers. While planning the architecture of buildings one might be interested in how people move around their intended design so that  shops, entrances, corridors, emergency exits and seating can be placed in useful locations. Pedestrian models are helpful in improving efficiency and safety in public places such as airport terminals, train stations, theatres and shopping malls. They are not only used as a tool for understanding pedestrian dynamics at public places, but also support transportation planners or managers to design timetables.

  A large number of models for pedestrian flow have appeared on different levels of description in recent years. The microscopic (individual-based) level models based on Newton type equations as well as vision-based models or cellular automata models and agent-based models have been developed, see Refs. \cite{he&Mo, helbing2, B01, group1,  deg, maury, piccoli}.  Hydrodynamic pedestrian flow equations involving equation for density and mean velocity of the flow are derived in Refs. \cite{bello, helbing, etikyala2014particle}. Modeling of pedestrian flow with scalar conservation laws coupled to the solution of the eikonal equation has been presented and investigated in Refs. \cite{ama, di, hughes, hughes1}. Pros and Cons of these models have been discussed in various reviews, we refer to \cite{bell,HJ10} for a detailed discussion of the different approaches.

 The modelling of pedestrian behaviour in a real-world environment is a complex problem.
 For example,  a majority of the people in a crowd are moving in groups and  social interactions  can greatly influence crowd behaviour.
 Most of the models mentioned above treat pedestrians as individual agents and neglect the group dynamics among them. The influence of group dynamics on the behaviour of pedestrians and the differences between the behaviour of pedestrians walking in groups and single pedestrians  have been presented in several recent works. We refer to 
 \cite{group1,group2,qui,SAD09,Koe14}.
 In these works  experimental studies as well as numerical experiments are presented

 In this work, we closely follow a procedure for interacting particle systems used, for example, in the description of coherent motion of animal groups such as schools of fish, flocks of birds or swarms of insects, see Ref. \cite{hyd,roosting}.
 It has been applied to pedestrian flow modelling in   Ref. \cite{etikyala2014particle}.
 There,   a classical microscopic social force model for pedestrians \cite{he&Mo} has been extended  with an optimal path computation as for example in Ref. \cite{hughes}.
 
One main objective of the present  paper is to include multi-group behaviour and the  impact of group dynamics, addressing in particular larger groups in a pedestrian crowd, into the set-up developped in \cite{etikyala2014particle}.
We extend the model developed there to the description of multi-group pedestrian flows using 
a multi-phase approach.
The dependence of the solutions on the level of attraction between the group members and the relaxation time towards the desired optimal path field is investigated and discussed.
 As a general result we observe an increase in evacuation time   by increasing the attraction between the group members.
 The second objective of the paper is to show the usefulness of using a unified approach for the numerical simulation of pedestrian models on microscopic and macroscopic scales. We use, as in Ref. \cite{etikyala2014particle},  particle methods on the microscopic and macroscopic level of the model hierarchy. These methods are straightforward for microscopic equations. In case of the macroscopic equations particle methods are based on a Lagrangian formulation of these equations and particles are used as grid points. A numerical comparison of different numerical approaches in microscopic and macroscopic situations is presented. 
 Moreover, we note that the method presented here  is  easily extended to more complicated "real" life situations, since the numerical implementation is   based on a  mesh-free fluid dynamic code for complex geometries.

 The paper is organized in the following way: in section 2 the hierarchy of multi-group pedestrian models is presented. Section 3 contains a description of the particle methods used in the simulations. Section 4 contains the numerical results. We consider an evacuation problem. A comparison of the solutions of the macroscopic equations is presented for  different parameters together with a comparison of the associated evacuation times. Finally, section 5 concludes the work.

\section{Multi-group pedestrian flow models}
In this section, we start with a multi-group microscopic model for pedestrian flow  using a microscopic social force model and a Hughes-type model including the solution of the eikonal equation.
We  proceed by  deriving  multi-group hydrodynamic and scalar models from the  microscopic model.
\subsection{The microscopic multi-group  model}
We consider a microscopic social force model for pedestrian flow including an optimal path computation. For references, see for example Refs. \cite{he&Mo,hughes}. For $N$ pedestrians divided into $M$ groups, we obtain a two-dimensional interacting particle system  with locations $x_i^{(k)} \in \mathbb{R}^2$, and velocity $v_i^{(k)} \in \mathbb{R}^2$. Here, the index  $i=1, \ldots N$ is used to number all pedestrians, the index $ k = 1,\ldots, M $ denotes the group to which the pedestrian belongs. $S^{(k)}$ 
 denotes the set of all $i$ which are in group $k$ and $N_k$ denotes the number of pedestrians in group $k$ with $ N = \sum_{l=1}^{M} N_l $.
 The equations of motion are
\begin{equation}\label{eq:2.1}
\begin{aligned}
    \frac{dx_i^{(k)}}{dt} &= v_i^{(k)} \quad\\
    \frac{dv_i^{(k)}}{dt} &= -  \sum_{l=1}^{M} \sum_{j \in S^{(l)}}   \nabla_{x_i} U^{(k,l)}(\mid x_i^{(k)} - x_j^{(l)}\mid) + G^{(k)}(x_i^{(k)},v_i^{(k)},\rho^N(x_i^{(k)})),
\end{aligned}
\end{equation}
where $U^{(k,l)}$ is an interaction potential denoting the interaction between members of groups $k$ and $l$. A common choice is the Morse potential
\begin{equation}\label{eq:2.2}
U^{(k,l)}(r) = - C_a e^{-r/l_a} + C_r e^{-r/l_r}.
\end{equation}
Here, $C_a$, $C_r$ are attractive and repulsive strengths and $l_a$, $l_r$ are their respective length scales. These constants depend on the groups $k$ and $l$ under consideration.
Similarly, one could use potentials given by poynomial or rational  functions.
An attractive interaction force acts only between  members of the same group. The repulsive force acts between all pedestrians. The acceleration towards the desired  direction  is given by
\begin{equation}
\label{eq:G}
 G^{(k)}(x,v,\rho^N) = \frac{1}{T} \left( - V^{(k)}(\rho^N)\frac{\nabla \Phi^{(k)}(x)}{\Vert\nabla \Phi^{(k)}(x)\Vert} - v\right).
 \end{equation}
Moreover,  $ \rho^N$ is given by 
$$ \rho^N(x) = \frac{1}{N}\sum_{l=1}^{M} \sum_{j \in S^{(l)}} \delta_S (x - x^{(l)}_j),$$
where $\delta_s$ is a smoothed version of the $\delta$-distribution with $ \int \delta_S (x)dx = 1 $ Finally, $\Phi^{(k)}$  is given by the solution of the  eikonal equation
$$ V^{(k)}(\rho^N(x))\Vert\nabla \Phi^{(k)}\Vert - 1 = 0 .$$
$V^{(k)}$ is a density-dependent velocity function, $V^{(k)}:\mathbb{R}^+ \longrightarrow \mathbb{R}^+$, $T$ denotes a reaction time. Moreover, we use the notation
$$\rho^{N,(l)}(x) = \frac{1}{N} \sum_{j\in S^{(l)}} \delta_S (x - x^{(l)}_j),$$
such that $ \rho^N(x) =  \sum_{l=1}^{M} \rho^{N,(l)}(x) $.

\begin{remark}
\label{proxemics}
The parameters in the above formulas, in particular in the definition of the interaction potential (\ref{eq:2.2}) should give average distances between the particles which are consistent
with  empirical data from social distance theory or proxemics, see \cite{H66,KGT09}. 
\end{remark}
\begin{remark}
In the definition of the acceleration towards the desired  direction (\ref{eq:G}), the speed with which the pedestrians are moving depends on the density around a pedestrian. In certain situations this could lead to unphysical effects, for example, if the pedestrian is approached from behind. A determination of the density including a "vision cone"  could be used here at the expense of a more complicated model.
\end{remark}
\begin{remark}
A further remark on the above microscopic model concerns the role of the interactions between the pedestrians.
Interactions are not only modelled by the interaction potential  $U$, but also by the Hughes type term (\ref{eq:G}).The motivation for the present way of modelling is a distinction between a  short scale 
interaction between the pedestrians in direct encounter described by the interaction potential $U$
and a  reaction of the pedestrian on a much larger spatial scale on the global density $\rho$  
 via the solution of the eikonal equation as in the Hughes approach.
In the present model, as in the Hughes model,  a knowledge of the density in the whole domain is assumed
for this second kind of interaction.
This could be changed to certain subregions of the computational domain by restricting the solution of the eikonal equation to these regions.
\end{remark}
\begin{remark}\label{centermass}
In \cite{group2} an attractive interaction of the members of the group with the 'center of mass' is postulated.
This gives an additional  term 
$$
\nabla_{x_i}  U_{CM}^{(k)} (\mid x_i^{(k)} - \frac{1}{N_k} \sum_{j\in S^{(k)}} x_j^{(k)}\mid).
$$
\end{remark}

\subsection{Mean-field equation}
Following \cite{braun1977vlasov,hyd,dobrushin1979vlasov,mean} one rescales the interaction potential of equation (\ref{eq:2.1}) with a factor $\frac{1}{N}$ and derives a kinetic mean field equation. This procedure is adapted to the multi-group case in the following. Our scaled microscopic model states
\begin{equation}\label{eq:2.3}
\begin{aligned}
    \frac{dx_i^{(k)}}{dt} &= v_i^{(k)} \quad\\
    \frac{dv_i^{(k)}}{dt} &= - \frac{1}{N}\sum_{l=1}^{M} \sum_{j \in S^{(l)}}   \nabla_{x_i}
    U^{(k,l)}(\mid x_i^{(k)} - x_j^{(l)}\mid) + G^{(k)}(x_i^{(k)},v_i^{(k)},\rho^N(x_i^{(k)})).
\end{aligned}
\end{equation}  

The empirical measures $f^{N,(k)}$ of the stochastic processes $z_i^{(k)}=(x_i^{(k)},v_i^{(k)})\in \mathbb{R}^2\times \mathbb{R}^2 $ are given by
$$
f^{N,(k)}(t,z)=\frac{1}{N}\sum_{i\in S^{(k)}} \delta(z-z_i^{(k)} ),
$$
where $\delta$ denotes the usual Dirac distribution and $z=(x,v)\in  \mathbb{R}^2\times \mathbb{R}^2$.
The mean field limit describes  the convergence for $N,N_k \rightarrow \infty$ of the  empirical measure $f^{N,(k)}$ towards the deterministic distribution $f^{(k)}$ of the stochastic process $z^{(k)}_t=(x^{(k)}_t,v_t^{(k)})$ governed by the so-called nonlinear McKean--Vlasov equation
\begin{equation}\label{kean}
\begin{aligned}
    \frac{dx^{(k)}}{dt} &= v^{(k)} \quad\\
    \frac{dv^{(k)}}{dt} &= - \sum_{l=1}^{M} 
      \nabla_{x} U^{(k,l)} \star \rho^{(l)} (\mid x^{(k)}   \mid)  
     + G^{(k)}(x^{(k)},v^{(k)},\delta_S \star \rho(x^{(k)})),
\end{aligned}
\end{equation}  
where $\rho^{(k)} (x) = \int f^{(k)}(x,v) dv $ and $\rho = \sum_{l=1}^{M} \rho^{(l)}$ and $\star$ denotes the convolution.
The corresponding differential equation for the evolution of the distribution functions $f^{(k)}= f^{(k)}(x,v,t)$ on state space, which is determined using It\^o's formula, is called the mean field equation. It is given by
 \begin{equation}\label{eq:2.4}
\partial_t f^{(k)} + v \cdot \nabla_x f^{(k)} + S^{(k)}f^{(k)} = 0
\end{equation}
with force term
 \begin{equation}
 \begin{aligned}
 S^{(k)}f^{(k)}(x,v)& =& \nabla_v \cdot \left(G^{(k)}(x,v,\delta_S \star \rho(x))f^{(k)}(x,v) \right)  \\
&&- \sum_{l=1}^{M}\nabla_v \cdot \left((\nabla_x U^{(k,l)} \star \rho^{(l)}) f^{(k)}(x,v)\right).
\end{aligned}
\end{equation}  
We note that due to our definitions we have 
 \[
\int \rho(x) dx = \int f(t,z)\, dz = 1 \quad \text{for all\; $t\ge 0$}.
 \]
For the following we define  the momentum by
$$\rho^{(k)}u^{(k)}(x) := \int v f^{(k)} (x,v) dv,$$
for $ k = 1, \ldots, M $.

\begin{remark} Let us mention that the rigorous passage from the microscopic particle system (\ref{eq:2.1})
towards the kinetic mean-field equation (\ref{eq:2.4}) as $N →\rightarrow \infty$ is a particular case of
the theory of well-posedness in measures for the kinetic equation (\ref{eq:2.4}), compare \cite{canizo2011well} for the classical case without a coupling to the eikonal equation. We refer also to the classical papers by \cite{dobrushin1979vlasov,braun1977vlasov} for convergence proofs in the deterministic case with a global Lipschitz condition on the  forces.
However, the present situation is more complicated due to the coupling of the microsocopic system to the eikonal equation. In this case, even the investigation of the limit equations is a non-trivial issues, see for example \cite{di} for the  one-dimensional Hughes equations.
We mention additionally, that a convergence proof of a particle system towards the Hughes equations has been investigated in \cite{di2} for a simple one-dimensional situation.
A rigorous proof of  the convergence of the above system (\ref{eq:2.1}) towards the mean field or the macroscopic equations is 
a completely  open issue.
\end{remark}

\begin{remark}
The additional  term 
$$
\nabla_{x_i}  U_{CM}^{(k)} (\mid x_i^{(k)} - \frac{1}{N_k} \sum_{j\in S^{(k)}} x_j^{(k)}\mid)
$$
gives in the limit
$$
\nabla_x U_{CM}^{(k)} ( \mid \int (x-y) \frac{\rho^{(k)} (y) }{\alpha^{(k)}}dy \mid )
$$
with
$$
\alpha^{(k)} =\int \rho^{(k)} (y) dy.
$$
\end{remark}

\subsection{The multi-group hydrodynamic model}
Hydrodynamic limits for interacting particle systems  have been derived for example in \cite{hyd, roosting}. We consider the mean field equation (\ref{eq:2.4}) and integrate the kinetic equation against $dv$ and $vdv$. Using  a mono-kinetic distribution function to close the resulting balance equations means that the velocity distribution function is assumed to be concentrated in the direction of the mean velocity, 
$$ f^{(k)} \sim \rho^{(k)}(x)\delta_{u^{(k)}(x)}(v).$$
Integrating the mean field equation with respect to  $dv$ one obtains the continuity equation for group $k$
\begin{equation}\label{eq:2.5}
\partial_t \rho^{(k)} + \nabla_x \cdot (\rho^{(k)} u^{(k)}) = 0.
\end{equation}
Integrating with respect to  $v dv$ yields the second balance equation  for group $k$
\begin{align}
\label{bal2}
\partial_t u^{(k)} + (u^{(k)} \cdot \nabla_x)u^{(k)} =\frac{1}{\rho^{(k)}} \int G^{(k)}(x,v,\delta_S \star \rho(x))f^{(k)}dv\\
 - \frac{1}{\rho^{(k)}} \int\sum_{l=1}^{M}\left( \nabla_x U^{(k,l)}\star\rho^{(l)}\right) f^{(k)}dv. \nonumber
\end{align}
Using now the mono-kinetic closure function we obtain
\begin{eqnarray}
\int G^{(k)}(x,v,\delta_S \star\rho(x))f^{(k)}(x,v)dv &=& \int G^{(k)}(x,v,\delta_S \star\rho(x))\rho^{(k)}(x)\delta_{u^{(k)}(x)}(v)dv \nonumber \\
&=& \rho^{(k)}(x)G^{(k)}(x,u^{(k)}(x),\delta_S \star\rho(x)) \nonumber
\end{eqnarray}
and
\begin{eqnarray}
\int\sum_{l=1}^{M}\nabla_x U^{(k,l)}\star\rho^{(l)}f^{(k)}(x,v)dv &=& \int\sum_{l=1}^{M}\left( \nabla_x U^{(k,l)}\star\rho^{(l)}\right) \rho^{(k)}(x)\delta_{u^{(k)}(x)}(v)dv \nonumber \\
&=& \rho^{(k)}(x)\sum_{l=1}^{M}\left( \nabla_x U^{(k,l)}\star\rho^{(l)}\right).\nonumber
\end{eqnarray}
Thus, equation (\ref{bal2}) becomes  the momentum equation 
\begin{equation}\label{eq:2.6}
\partial_t u^{(k)} + (u^{(k)} \cdot \nabla_x)u^{(k)} = G^{(k)}(x,u^{(k)}(x),\delta_S \star\rho(x)) - \sum_{l=1}^{M}\left( \nabla_x U^{(k,l)}\star\rho^{(l)}\right) 
\end{equation}
with 
\begin{equation}\label{eq:2.7}
 G^{(k)}(x,u^{(k)},\delta_S \star\rho(x)) = \frac{1}{T} \left( - V^{(k)}(\delta_S \star\rho(x))\frac{\nabla \Phi^{(k)}(x)}{\Vert\nabla \Phi^{(k)}(x)\Vert} - u^{(k)}\right),
\end{equation}
for $ k = 1, \ldots, M $ and this is coupled to
$$ V^{(k)}(\delta_S \star\rho(x))\Vert \nabla\Phi^{(k)}(x)\Vert = 1 .$$

\subsection{The multi-group scalar model}
In this section, we reduce the hydrodynamic description deriving scalar models and connect the approach to the Hughes model. We assume again an interaction potential depending only on $x$. Starting from the hydrodynamic momentum equation derived from the standard Maxwellian closure we neglect time changes in this equation and obtain an equation for $u^{(k)}$ as
$$ G(x,u^{(k)},\delta_S \star\rho(x)) = \sum_{l=1}^{M}\nabla_x U^{(k,l)}\star\rho^{(l)}.$$
Using equation (\ref{eq:2.7}), we get
$$ u^{(k)} = - \sum_{l=1}^{M}T \nabla_x U^{(k,l)}\star\rho^{(l)} - V^{(k)}(\delta_S \star\rho(x))\frac{\nabla \Phi^{(k)}(x)}{\Vert\nabla \Phi^{(k)}
(x)\Vert}. $$
Thus, the resulting scalar equation for $\rho^{(k)}$ is
\begin{equation}\label{eq:2.8}
\partial_t \rho^{(k)} + \nabla_x \cdot \left[ \rho^{(k)}\left(- \sum_{l=1}^{M}T \nabla_x U^{(k,l)}\star\rho^{(l)} - V^{(k)}(\delta_S \star\rho(x))\frac{\nabla \Phi^{(k)}(x)}{\Vert\nabla \Phi^{(k)}(x)\Vert}\right) \right]  = 0,
\end{equation}
for $ k = 1, \ldots, M $.

A further simplification is obtained approximating the potential $ U^{(k,l)} $ by a $ \delta $ distribution, i.e.,
$$U^{(k,l)}(y) \sim D^{(k,l)}\delta_0(y)$$
with the constant $ D^{(k,l)} > 0 $ given by
$$D^{(k,l)} = \int U^{(k,l)}(y)dy. $$
Moreover,
$$  \delta_S \sim \delta.
$$
This yields straightforwardly
\begin{eqnarray}
\nabla_x U^{(k,l)}\star\rho^{(l)} = D^{(k,l)} \nabla_x \rho^{(l)}. \nonumber
\end{eqnarray}
Hence, equation (\ref{eq:2.8}) becomes a multi-group version of the Hughes equations
\begin{equation}\label{eq:2.9}
\partial_t \rho^{(k)} - \nabla_x \left( V^{(k)}(\rho(x))\frac{\nabla \Phi^{(k)}(x)}{\Vert\nabla \Phi^{(k)}(x)\Vert}\rho^{(k)}\right) = \sum_{l=1}^{M} D^{(k,l)}T\nabla_x\left( \rho^{(k)}\nabla_x \rho^{(l)}\right) ,
\end{equation}
where $ k = 1, \ldots, M $. This  is again combined with the Eikonal equation
$$V^{(k)}(\rho(x)) \Vert\nabla\Phi^{(k)}(x)\Vert = 1.$$

\begin{remark}
\label{discussion}
Looking at equation (\ref{eq:2.9}) one observes that the influence of the diffusive term can be decreased in different ways. For example, adding or increasing the attraction interaction will decrease the value of $D^{(k,l)}$. Decreasing the value of the relaxation time $T$ will also decrease the diffusion.
From considering (\ref{eq:2.9}) together with a monotone decaying function 
$V^{(k)}$ one would expect an increase of the diffusion to lead to a decay of the maximal values of the density and then to a faster transport in the direction of the eikonal field, at least for simple flow situations.
\end{remark}

\section{Numerical methods}
In this section, we discuss the numerical methods for the two  multi-group models   (\ref{eq:2.5}), (\ref{eq:2.6}) and (\ref{eq:2.8}).

\subsection{General framework for particle methods for the hydrodynamic model}
To solve the hydrodynamic limit equations  numerically we use a  particle method, see, for example \cite{fpm}. Mesh-less or particle methods are an appropriate way to solve pedestrian flow problems due to the appearance of situations with complicated geometries, free and moving boundaries and potentially large deformations of the domain of computation, i.e. the region where the density of pedestrians is non-zero. The particle method is based on a Lagrangian formulation of the hydrodynamic equations (\ref{eq:2.5}) and (\ref{eq:2.6}). We consider
\begin{eqnarray}
\frac{dx^{(k)}}{dt} &=& u^{(k)} \nonumber \\
\frac{d\rho^{(k)}}{dt} &=& -\rho^{(k)} \nabla_x \cdot u^{(k)} \nonumber \\
\frac{du^{(k)}}{dt} &=& G^{(k)}(x,u^{(k)},\delta_S\star\rho) - \sum_{l=1}^{M} \nabla_x U^{(k,l)}\star \rho^{(l)} ,\nonumber
\end{eqnarray}
where  $\frac{d}{dt} = \partial_t + u^{(k)} \cdot \nabla_x $ and $k = 1, \ldots, M$.

Meshfree Lagrangian methods use for the quantities appearing on
the right hand side of the above  equations a difference approximation at the particle locations
from the surrounding neighbouring particles using weight functions and a least
square approximation. For the present computation we use weight functions $w$ with compact support of radius h,
restricting in this way the number of neighbouring particles. 
The Gaussian weight functions are of the form
\begin{eqnarray} 
w = w(  x; h) =
\left\{ 
\begin{array}{l}  
\exp (- \alpha \frac{ \vert x  \vert^2 }{h^2} ), 
\quad \mbox{if    }  \frac{|  x  |}{h} \le 1 
\\
 0,  \qquad \qquad \quad \quad \quad \mbox{else}
\end{array}
\right.
\label{weight}
\end{eqnarray}
The radius h is chosen to
include initially enough particles for a stable approximation of the equations, which
is approximately three times the initial spacing of the particles. During the computation
a particle management has to be implemented, such that particles are added
or removed in case the local distribution of the particles becomes too rarefied or too
dense, respectively. See \cite{tiwari} for details of the implementation.

 The simplest way to evaluate the integral over the interaction potential is to use  a straightforward first order integration rule using an approximation of the local area around a particle determined by nearest neighbour search. This works fine for a well resolved situation with a sufficiently large number of gridpoints. 
 The resulting system of ODEs is then solved  by a suitable  time discretization method. 
The above considerations show that, if the number of macroscopic gridpoints is   approximately equal to the (large) number of microscopic particles, then  the macroscopic computations are essentially equivalent  to a microscopic solution of equation (\ref{eq:2.3}).
 
 The reconstruction of the density out of the particle locations is done by  using for $\delta_S$ again functions with compact support. Here we have chosen a quadratic 
 polynomial and the radius  $R$ and  a normalization factor $C_R$ to define
 $$
 \delta_S (x) = C_R (R-\vert x\vert)^2 ,$$
 for $\vert x \vert \le R$ and $0$ otherwise.   
 
Boundary conditions are realized by using fixed boundary particles with a suitable interaction potential. In the present  case we use a purely repelling quadratic potential.
 We note that the time step of the computation has to be adapted to the strength of the boundary potential in order to obtain a stable method.

\subsection{A multi-scale approach based on the mean field approximation}

A situation as described above  with a  number of 
macroscopic gridpoints  approximately equal to  the number of  pedestrians does not require a special algorithm including any mean-field or macroscopic considerations.
However,  if the number of 'real' microscopic particles is very large, that does not mean that the number of macroscopic grid-particles in the particle method has to be increased in the same way, since
the grid particles only 
 play the role of discretization points.  
 The key point of the method, compare \cite{KT14}, is to approximate the convolution integrals
 appearing in the above equations not by a simple Riemann sum, which would essentially lead to a microscopic computation for equation (\ref{eq:2.3}), but by a higher order approximation of the functions on the respective Voronoi cells, compare again \cite{KT14}.
 This approach  yields in the macroscopic limit, where  $U$ and $\delta_S$ can be approximated by a $\delta$ function,  an accurate  method for the limiting macroscopic equations (\ref{eq:2.9}) or its hydrodynamic counterpart,
 even if the number of macroscopic grid particles is still small compared to the real microscopic number of pedestrians.

This approach allows to use in certain situations a much smaller amount of particles. In turn, this  reduces  the numerical efforty considerably, which is 
 essentially determined by the number of particles in the computation.
We refer to \cite{KT14,KT17} for a thorough discussion of this issue and of the multi-scale numerical algorithm.

We note, that in this way,  the numerical method for the hydrodynamic equations ranges from a 
"nearly" microscopic " solver to a purely macroscopic solver depending on the number of
grid-particles involved in the computation compared to the "real" number of physical particles.

\begin{remark}
 The scalar equation is solved  with a similar particle method. In this case the so called diffusion velocity methods is used, i.e., the equation (\ref{eq:2.8}) is written as a pure transport problem
\begin{equation}\label{eq:3.10}
\partial_t \rho^{(k)} + \nabla_x \cdot \left( a \rho^{(k)}\right) = 0,
\end{equation}
with 
$$a = - \sum_{l=1}^{M}T \left( \nabla_x U^{(k,l)}\star\rho^{(l)}\right)  - V^{(k)}(\delta_S \star \rho(x))\frac{\nabla \Phi^{(k)}(x)}{\Vert\nabla \Phi^{(k)}(x)\Vert} $$
and then solved in a Lagrangian way. The approximation of the convolution term and the realization of the boundary conditions is done as for the hydrodynamic models. 
\end{remark}

\begin{remark}
In all cases the solution of the eikonal equation is coupled to the flow simulation. The eikonal equation is solved by a fast marching method, see Ref. \cite{eikonal1}. We use a similar methodology as described in Ref. \cite{eikonal1} to solve the eikonal equation on an unstructured fixed grid. Interpolation beween the grids uses a least squares method.  We update the eikonal solution in every tenth time step in order to save computational time.
\end{remark}

\begin{remark}
Finite Volume methods. The macroscopic  equations (\ref{eq:2.5}),(\ref{eq:2.6}) could also be solved with a Finite Volume method. However, in the present context a particle method is more natural, since,  due to the Lagrangian formulation, one obtains an accordance with   a microscopic approach in the limit of a fine resolved situation with a large number of grid points. Moreover, situations with free surfaces as in the example below
are more easily and naturally treated in a Lagrangian particle method than in a classical FV method.
\end{remark}

\section{Numerical results}
\begin{figure}[htbp]
\centering
  \includegraphics[width=.45\linewidth]{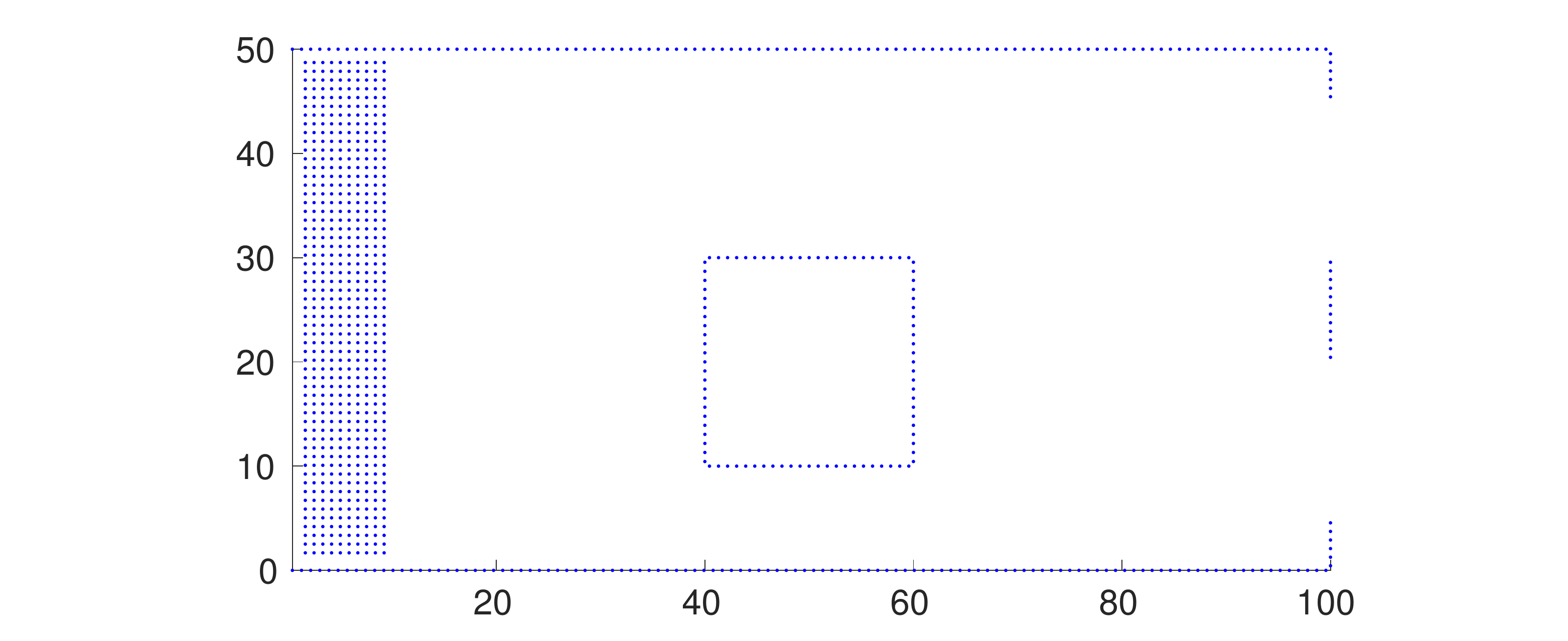}
 \caption{Geometry}
\label{fig:1}
\end{figure}

In this section, we present a series of numerical experiments for single and  multi-group hydrodynamic (\ref{eq:2.5}), (\ref{eq:2.6})  equations applied to an evacuation problem. We investigate the model numerically for a configuration defined in Ref. \cite{num}. All distance are measured in meter $m$. Densities are measured in $1/m^2$. Time is measured in seconds $s$ and velocity in $m/s$. Consider a railway platform of  length $100$
and  width $50$ with a square obstacle of size $20  \times 20  \; $ centered around the point $(50 , 20)$.
Rescaling space and time with a parameter $\alpha$ and varying $\alpha$ allows to consider more "microscopic" or more  "macroscopic" situations.
Initially, pedestrians are concentrated at the left boundary.
They can leave  at  either of  two exits of  width $15$ on the right boundary.
The initial pedestrians are concentrated in a region $[l  ,r ] \times [0,50 ]$. Having, for example,  an initial  density of one pedestrian per square meter, $\rho = 1 $,
and $l=0, r=10$,  see Figure \ref{fig:1},  we  interpret all spatial distance $x$ as $ \alpha x \; $ and
 obtain a total number of $500  \; \alpha^2$ pedestrians initially.

We choose the inflow velocity as  $ (V^{(k)}(\rho),0) $, $k=1,2$ , where the speed-density relation, $V^{(k)}(\rho) = V_{max}(1-\rho/\rho_{max}) $    if $\rho \le \rho_{max} $ and $0 $ otherwise.
Here, $\rho_{max} = 10$
and   $ V_{max} = 2  $ has been chosen. Moreover, we choose  if not otherwise stated  $T = 0.001 $. We are choosing different values for the  attractive strength $ C_a $, repulsive strength $ C_r $, attractive interaction length $ l_a $, and repulsive length $ l_r $. We use an explicit time integration for solving the hydrodynamic  model with the constant time step $ \Delta t = 0.00042 $ for all experiments.

In the following we investigate different issues. First, we give  results on the 
behaviour of the numerical algorithm including a numerical comparison of the multi-scale solver based on the 
 mean-field approach described above  and a  microscopic solver for single and multi-group pedestrian flow
 based on equation (\ref{eq:2.3}).  

Second, we investigate the influence of the different parameters in the multigroup model.
Single- and multi-group models are compared to each other and, more general, situations with increasing attraction between the members of the group are investigated. 
Finally, we discuss the results obtained here and compare them to experimental data and  numerical results
available in the literature.

\subsection{Comparison of numerical algorithms}

\subsubsection{Comparison of  numerical algorithms for single phase flow}

In this subsection we consider   a single phase  pedestrian flow. For the single group pedestrian model the  interaction between the pedestrians is given by a purely   repulsive interaction potential.
For simplicity we choose here a quadratic polynomial $U(r) = C  (2 R -\vert x \vert)^2$ restricted to a circular region with radius $R$ around the pedestrian. $R$ is chosen equal to $0.4$ and $C=1000$.
The hydrodynamic macroscopic equations (\ref{eq:2.5}) and (\ref{eq:2.6}) are discretized  with different numbers of grid points and
 the quality of the macroscopic approximation of the microscopic problem is investigated.
  The hydrodynamic equations are solved using the multiscale approach, for detail we refer to \cite{KT14,KT17}. For comparison we  also show   the results of microscopic computations with smaller numbers of particles.
  
The computational domain and the boundary conditions are same as in the earlier cases.
The initial value is chosen as  $\rho = 2$ in the region $[0, 30] \times [0, 50]$.   We consider the above described situation with $\alpha =\sqrt{\frac{37500}{3000}}\sim 3.5$. That means we have
initially $37500$ pedestrians. This is equivalent to an initial distance of gridpoints equal to $\Delta x = 0.2$ in unscaled coordinates.
Moreover, the scaling yields a physical interaction radius equal to $3.5 \times 0.4 = 1.4 $. 
The microscopic simulation is now compared with a macroscopic simulation with different numbers of gridpoints.
We vary the initial average distance between grid points from 0.2 to 1, i.e. the number of grid particles varies between $1500$ and $37500$.  

\begin{figure}
	\centering
	\captionsetup[subfigure]{margin=5pt} 
	\subfloat[Microscopic, $\Delta x =1$]{
		\includegraphics[keepaspectratio=true,  width=.48\textwidth]{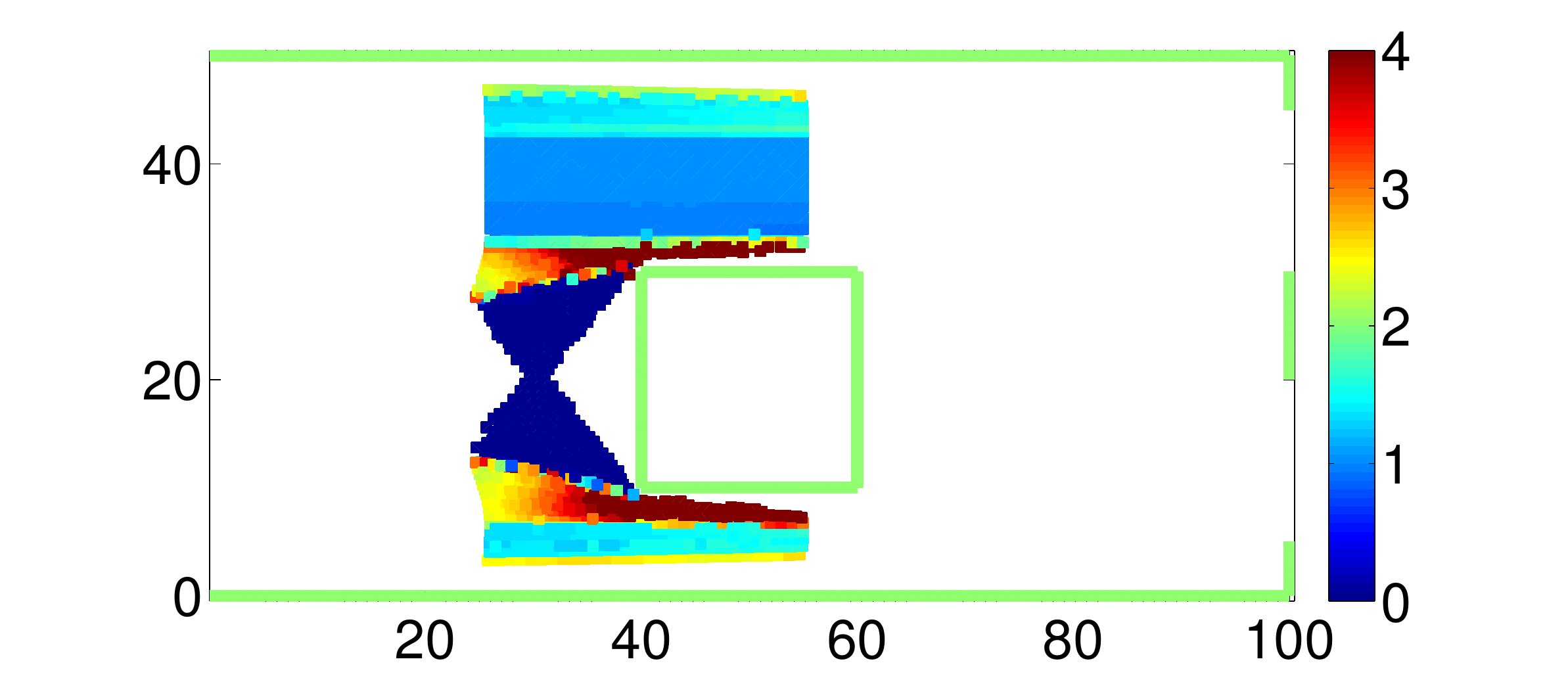}
	} \subfloat[Multi-Scale,  $\Delta x = 1$]{
		\includegraphics[keepaspectratio=true, width=.48\textwidth]{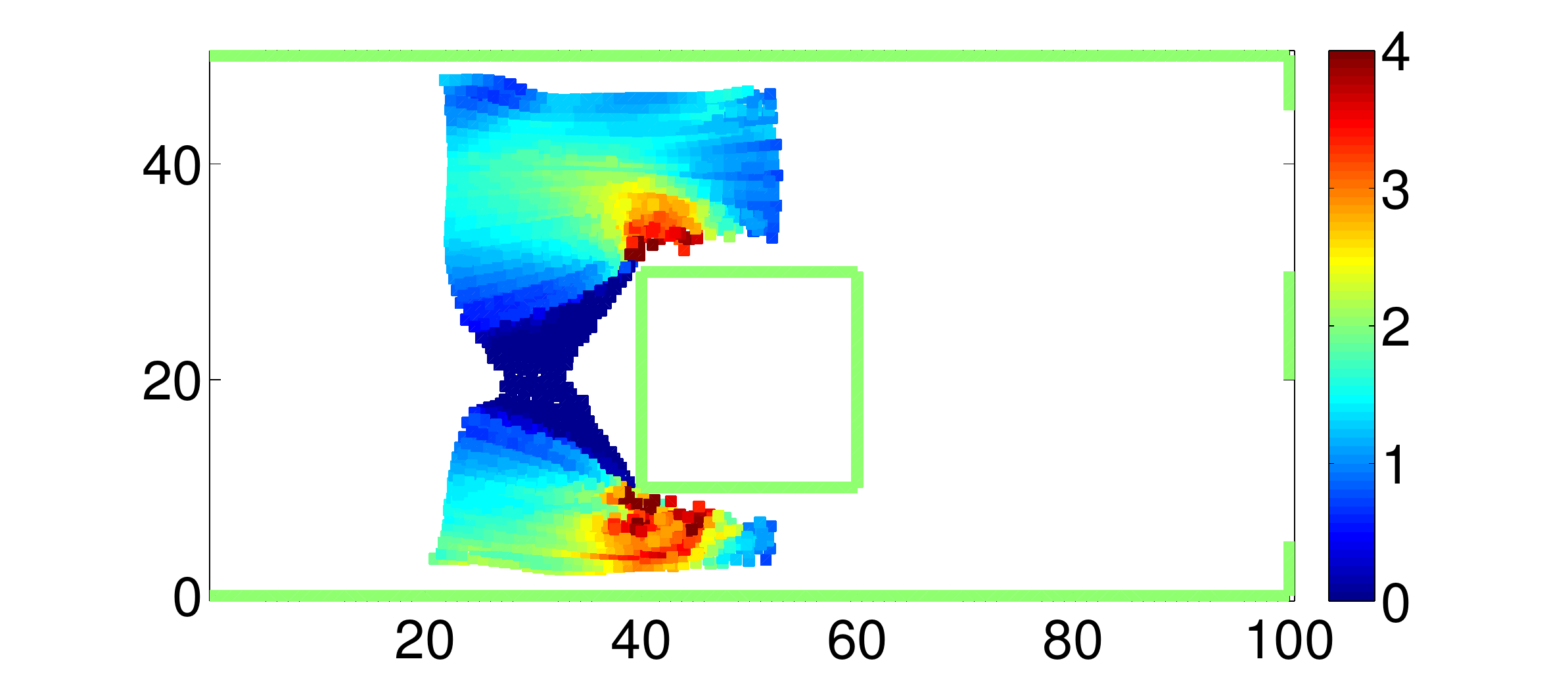}
	} 
	\captionsetup[subfigure]{margin=5pt} 
	\subfloat[Microscopic, $\Delta x =0.2$]{
		\includegraphics[keepaspectratio=true, width=.48\textwidth]{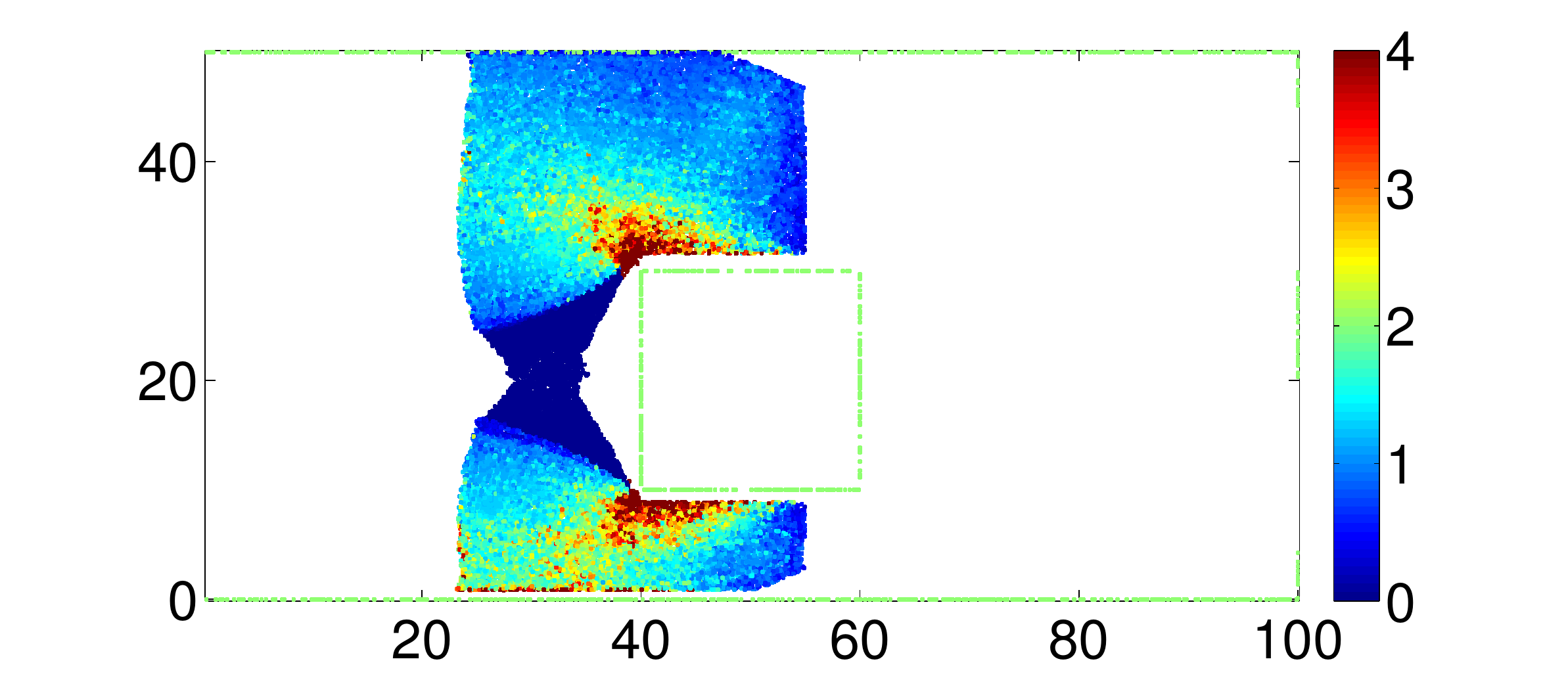}
	} \subfloat[Multi-scale, $\Delta x =0.2$]{
		\includegraphics[keepaspectratio=true, width=.48\textwidth]{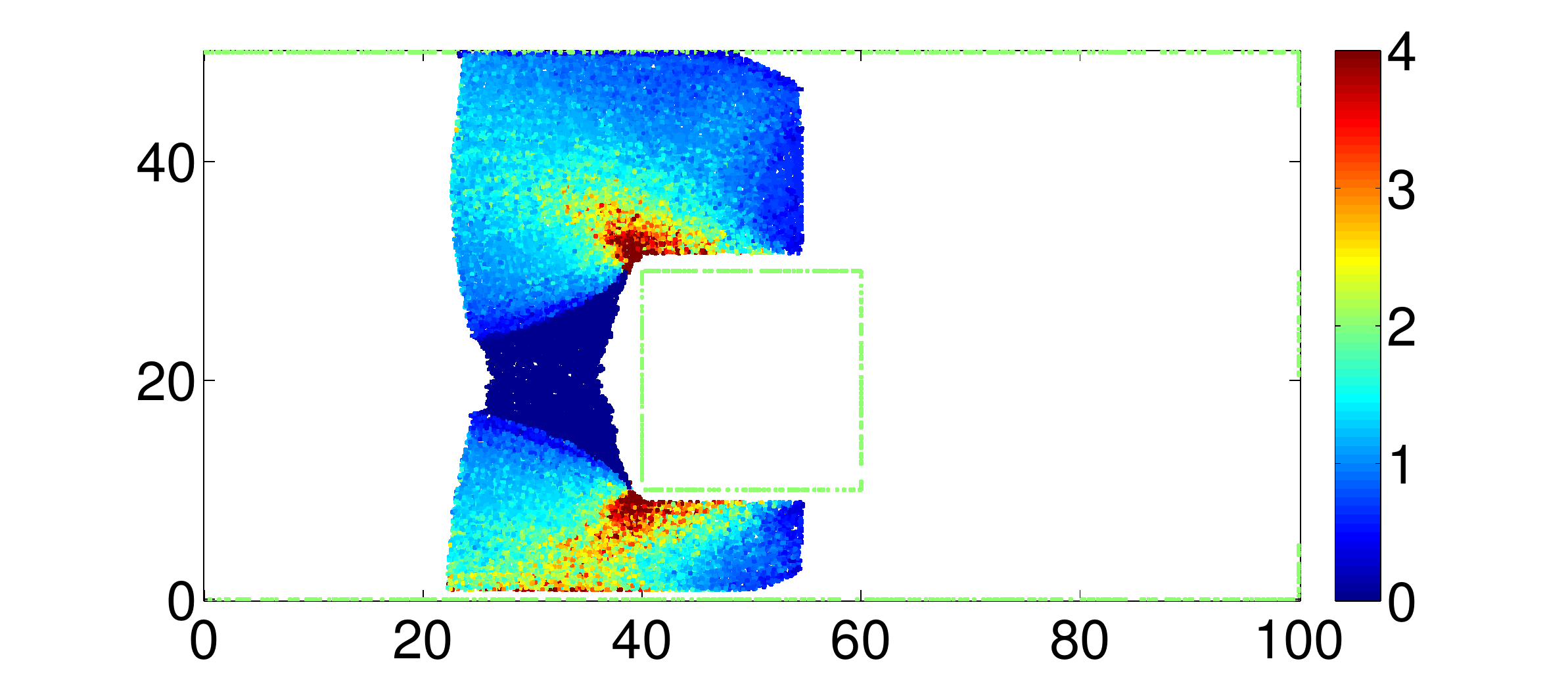}
	}
	\caption{Density plot determined from  with microscopic  and multiscale  approximations at time $t = 12.5$ for $\Delta x = 1$ , $\Delta x = 0.2$.   }
	\label{fig:10}
\end{figure}

In Figure \ref{fig:10} we show a comparison of solutions obtained from the microscopic and macroscopic multiscale method. Figure \ref{fig:10} (c) shows the result of a microscopic simulation with 37500 particles.
Using a grid with the same number of particles we obtain a very similar result, see Figure \ref{fig:10} (d).

Figure \ref{fig:10} (b) shows the result of the multi-scale approach with only 1500 grid points.
One obtains still a reasonable approximation of the physical situation. This is in contrast to a microscopic approximation, where just 1500 pedestrians are used, see  \ref{fig:10} (a), yielding a completely different flow pattern.

\begin{figure}
	\centering
	\includegraphics[keepaspectratio=true,   width= .65\textwidth]{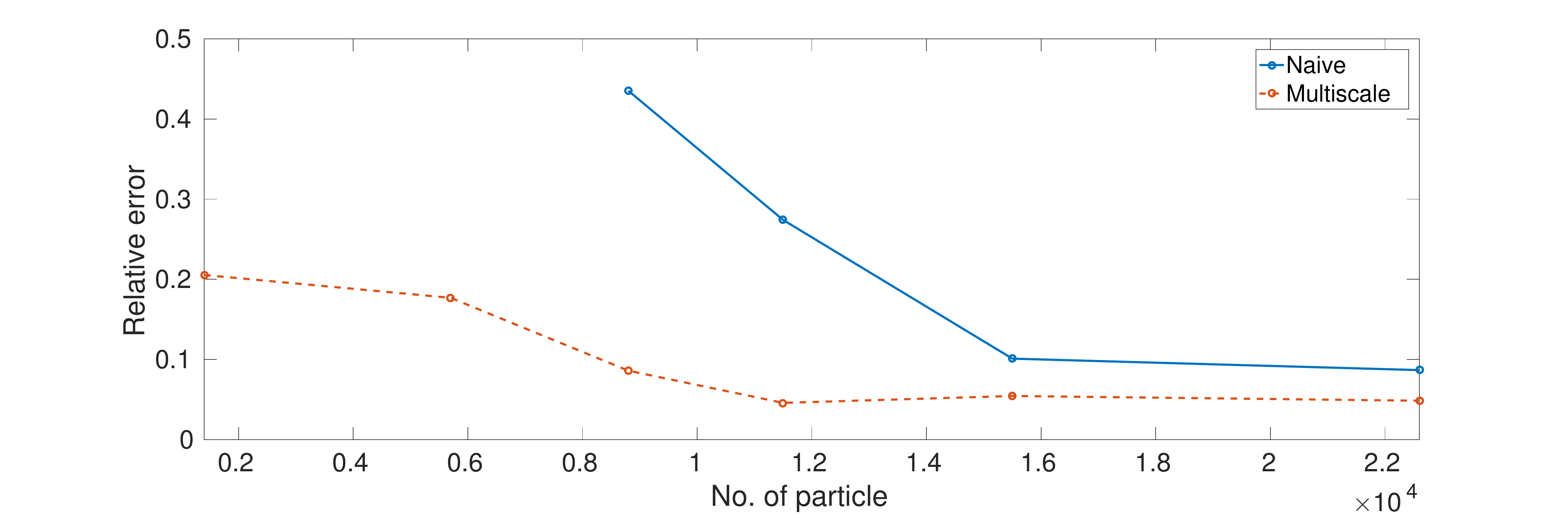}  
	\caption{Error plot for microscopic and multiscale simulations.}
	\label{fig:11}
\end{figure}

A more quantitative comparison gives the following.
In Table \ref{table1} and Figure \ref{fig:11}
we compute the differences to the microscopic reference solution with $37500$ particles and the CPU times of microscopic and multi-scale  method for different numbers of grid-particles. The differences  are determined (in unscaled coordinates) along a line with $y=37$ and $x \in [25,50]$. The relative $\mathcal{L}^2$-errors  are given  as well as the computation times in minutes.
The microscopic reference solution is computed by  using a spacing of $\Delta x = 0.2$ and
approximately $37500$ particles. The difference between this solution and the  solution of the multi-scale hydrodynamic method with the same number of grid points is   
of the order  $10^{-2}$.
The solution determined from a microscopic simulation with $N=1500$ and $N=6000$ shows a completely different flow pattern. Therefore no values for the error are given in Table \ref{table1}.
Looking at the CPU times in Table \ref{table1}   one observes  a large gain in computation time using  the  multi-scale simulation can be obtained for a very small loss in accuracy.

\begin{table}
	\begin{center}
		\begin{tabular}{|r|r|r|r|r|}
			\hline
			initial   &$\# $ particles  & microscopic  &multi-scale & CPU time  \\ 
			spacing &$ $ &error & error  &  \\ \hline 
			$ 1$ & $1500$ &- & $0.20 $  & $10$ min   \\
			$0.5$ & $6000$ &- & $0.17 $  & $28$ min   \\
			$0.4$ & $9375$ &$0.44$ & $0.09 $  & $43$ min   \\
			$0.35$ &$12245$ &$ 0.27$ &$0.05 $   & $62$ min      \\ 
			$0.3$ &$16666$ &$ 0.1$ &$0.05 $   & $87$ min      \\ 
			$0.25$ &$24000$ &$ 0.09$ &$0.05 $   & $163$ min      \\ 
			$0.2$ & $37500$ & -& -  & $233$ min  \\   \hline
		\end{tabular}
		\caption{Comparison of CPU times between microscopic and multiscale simulations. The error analysis is performed at time $t=12.5$  }
		\label{table1}
	\end{center}
\end{table}

\subsubsection{Comparison of multigroup  micro and multi-scale  simulations}

\begin{figure}[htbp]
\centering
\subfloat[Multi-group pedestrian]
  {\includegraphics[width=.45\linewidth]{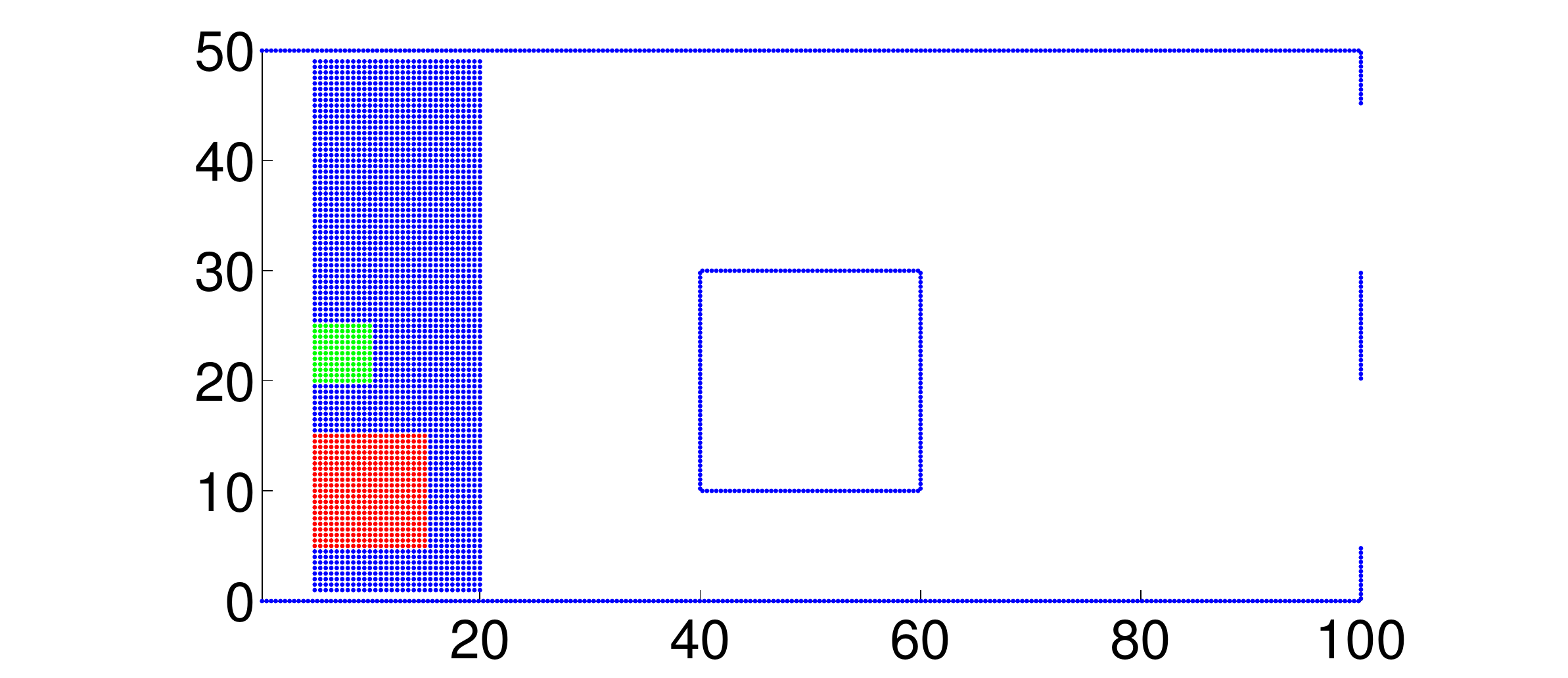}}
 \caption{Initial configuration.}
\label{fig:initial2}
\end{figure}

In this subsection we consider  a multi-group pedestrian flow  with initial data shown in Figure \ref{fig:initial2}. 
The multi-group pedestrian model contains three groups: the first larger one consists of pedestrians interacting with each other with a  purely repulsive interaction term as in
the single group model. The second and third group consist of  
pedestrians with an additional attraction between the members of the respective  groups.
We have chosen here the interaction potential given by equation (\ref{eq:2.2}) with
$l_r =0.5, l_a =1, C_r =200$ and $C_a = 50$. 
 For the single pedestrians we use $C_ a=0$.
 A value of $l_r =0.5$ gives a relevant repulsive force for distances smaller than $1$.
In this case the pedestrian are initially located in $[5,20]\times[0, 50]$, where one group is located in  $[5, 15]\times[5, 15]$ and the second one is located in $[5, 10]\times[20, 25]$.
See Figure  \ref{fig:initial2} for the location of the two groups.
The density is initially $ \rho = 2 $.

\begin{figure}
	\captionsetup[subfigure]{margin=5pt} 
	\subfloat[$\Delta x = 1, Micro $]{
		\includegraphics[keepaspectratio=true, width=.48\textwidth]{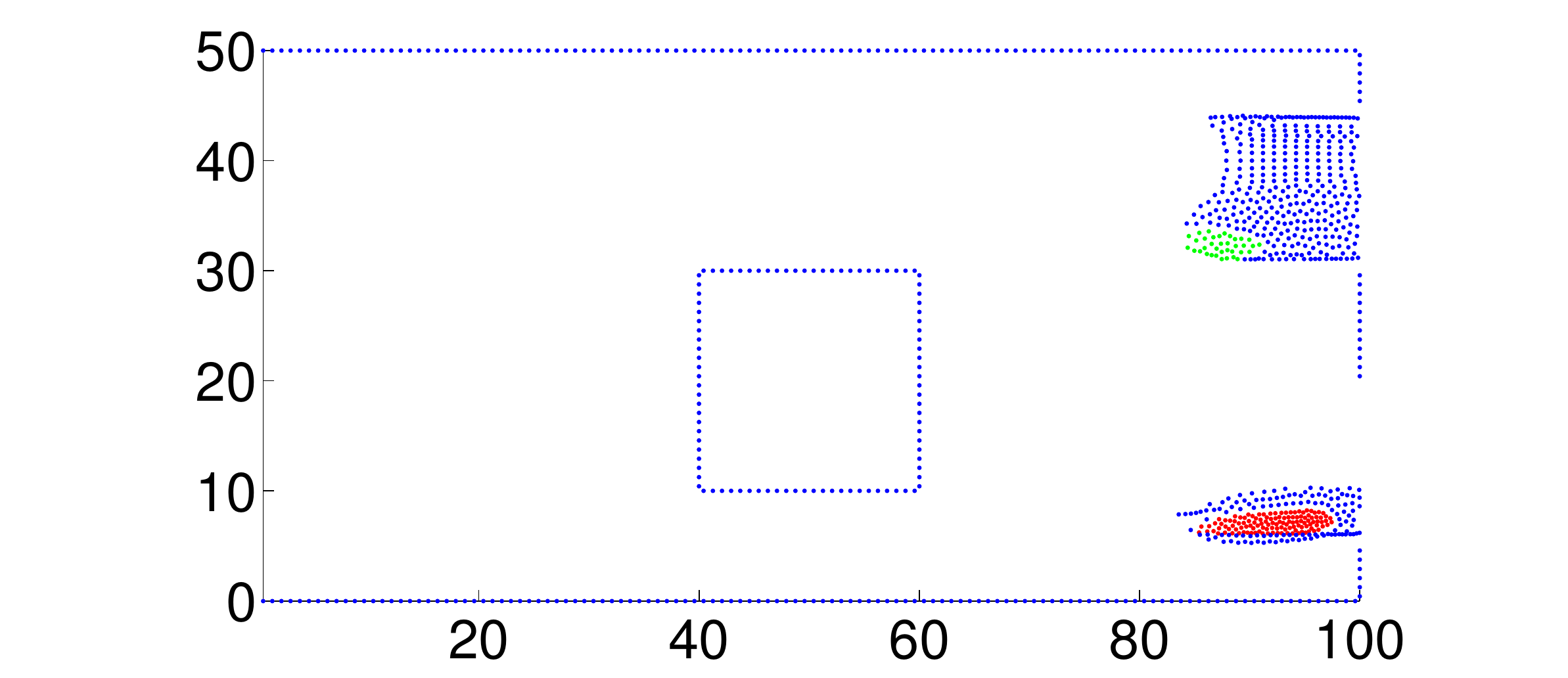}
	} \subfloat[ $\Delta x = 1$, Multiscale]{
		\includegraphics[keepaspectratio=true, width=.48\textwidth]{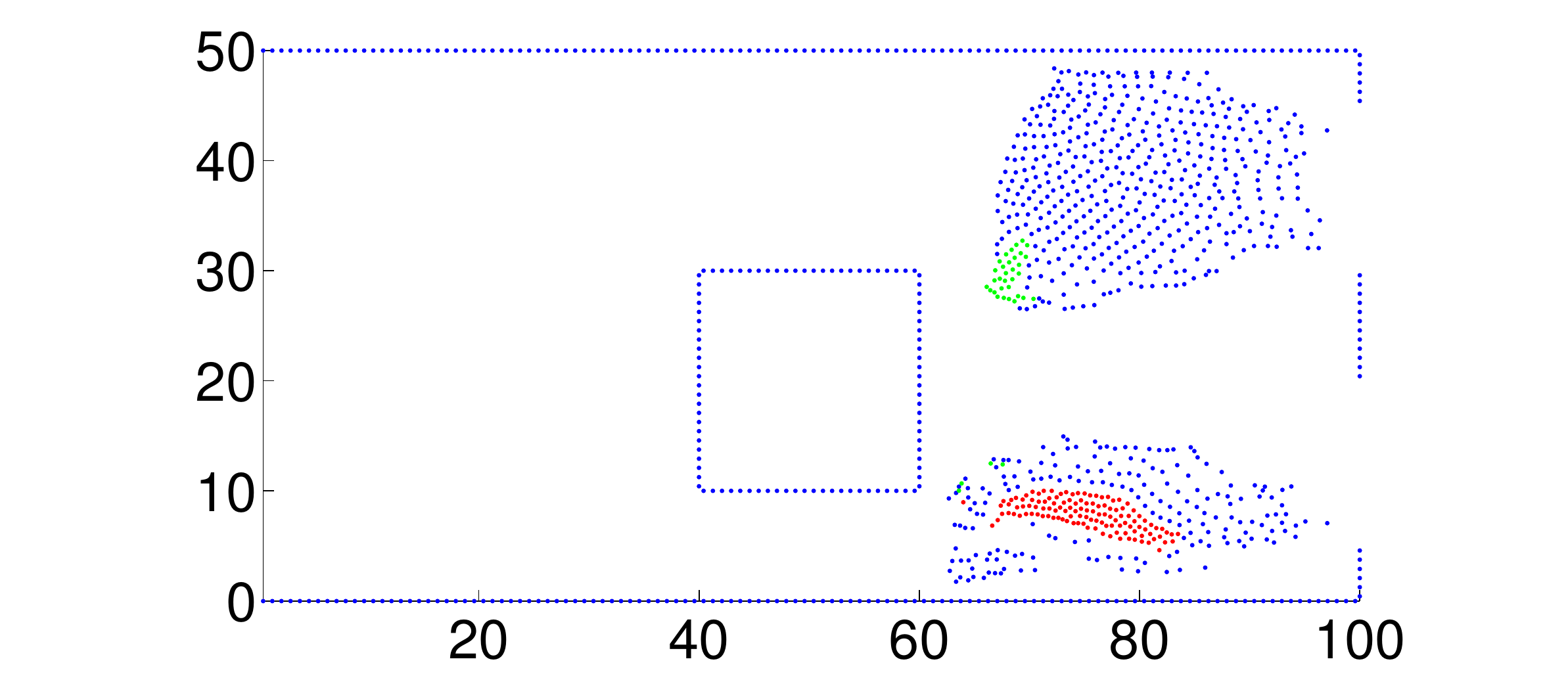}
	} 
	\captionsetup[subfigure]{margin=0pt} 
	\subfloat[ $\Delta x = 0.5$, Micro]{
		\includegraphics[keepaspectratio=true, width=.48\textwidth]{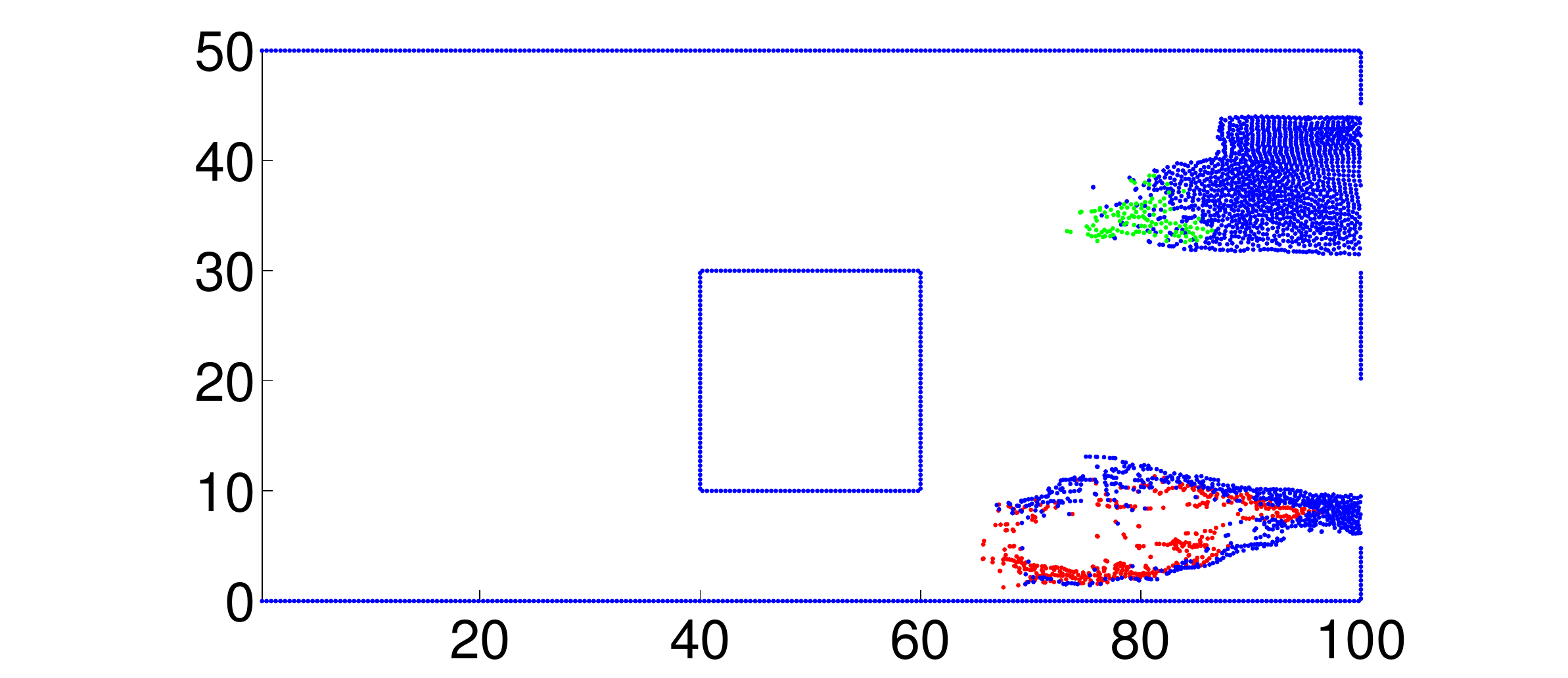}
	}
	\subfloat[$\Delta x = 0.5$, Multiscale ]{
		\includegraphics[keepaspectratio=true, width=.48\textwidth]{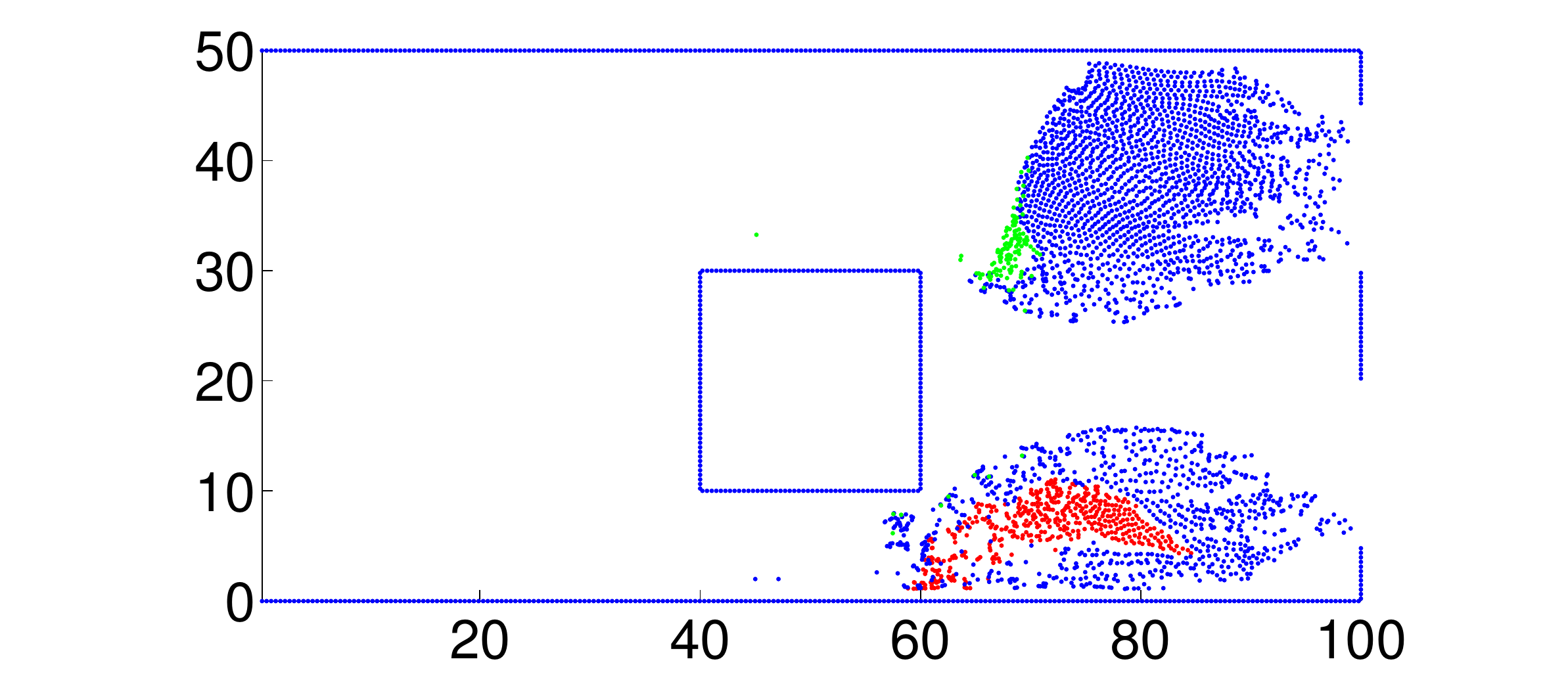}
	}   \\
	\subfloat[ $\Delta x = 0.25$, Micro]{
		\includegraphics[keepaspectratio=true, width=.48\textwidth]{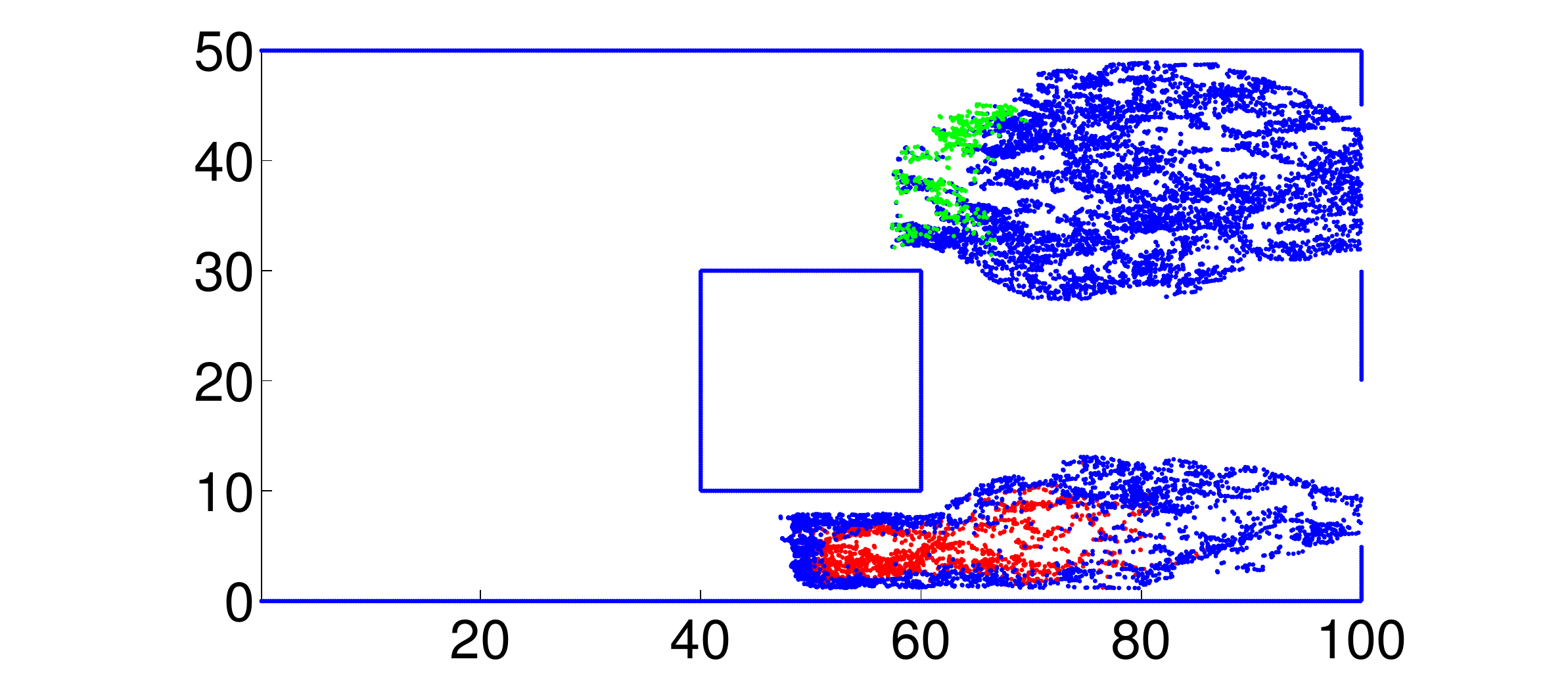}}   
	\subfloat[ $\Delta x = 0.25 $, Multiscale]{
		\includegraphics[keepaspectratio=true, width=.48\textwidth]{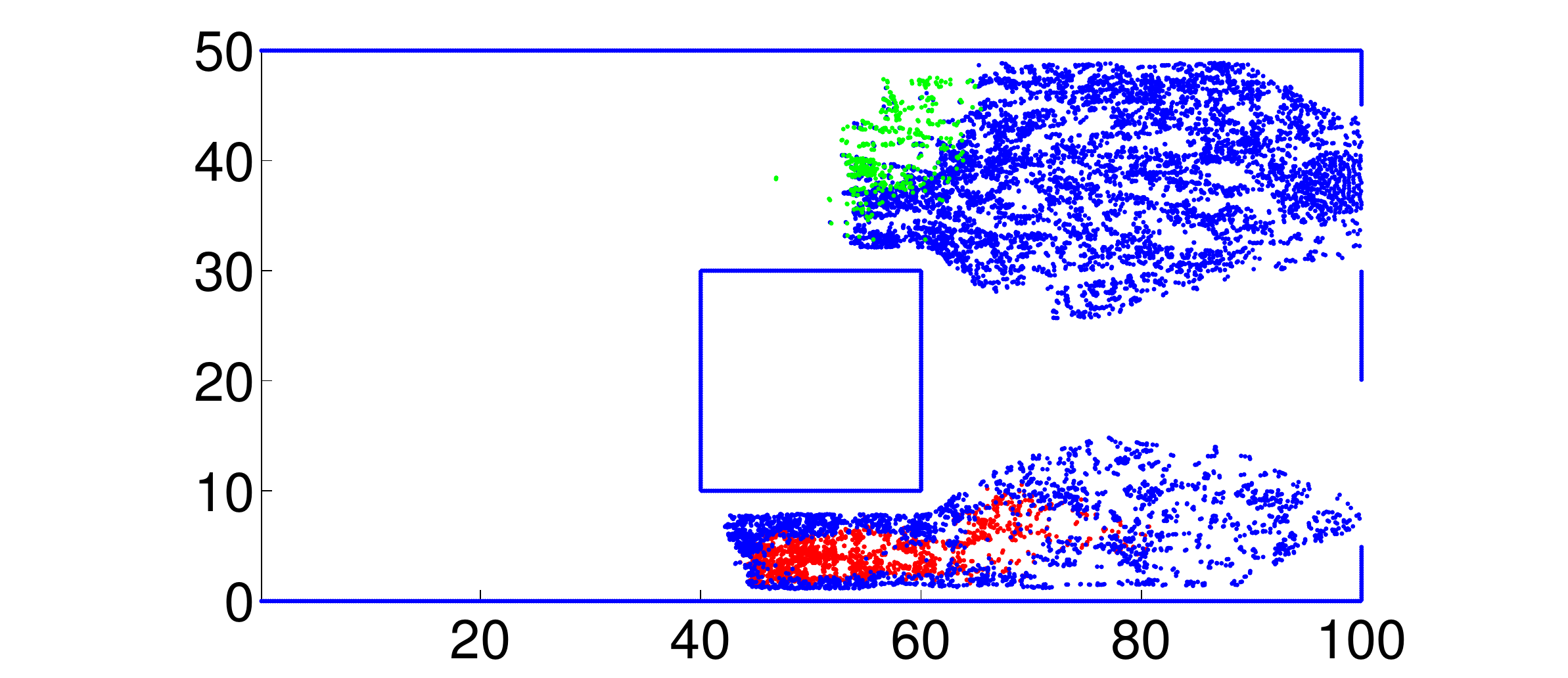}}  
	\captionsetup{margin=20pt}
	\caption{The position of group and individual pedestrians obtained from microscopic and multiscale approaches at time $t=40$  for initial particle spacing $\Delta x$ ranging from $1$ to $0.25$.  }
	\label{fig:multi}
\end{figure}

We consider the above situation with $\alpha =\sqrt{\frac{12000}{1500}} \sim 2.8$. That means we have
$12000$ pedestrians. Thus, the physical distance for repulsion is  again in a reasonable range of values.
Furthermore, we have simulated the multigroup pedestrian flow model with microscopic and multiscale algorithms.  In Figure \ref{fig:multi} we have plotted the 
positions pedestrians obtained from the microscopic and multiscale simulations  at time $t = 40$  with initial spacing $\Delta x $ equal to $1, 0.5$ and $0.25$, which corresponds approximately  to the number of particle $N=750, 3000$ and $12000$, respectively. We observe that the structure of the multiscale solutions  is even  for smaller numbers of particles similar to the microscopic solution for $12000$ pedestrian. 

\subsection{Comparison of modelling approaches}

\begin{figure}[htbp]
\centering
\subfloat[Multi-group pedestrian]
  {\includegraphics[width=.45\linewidth]{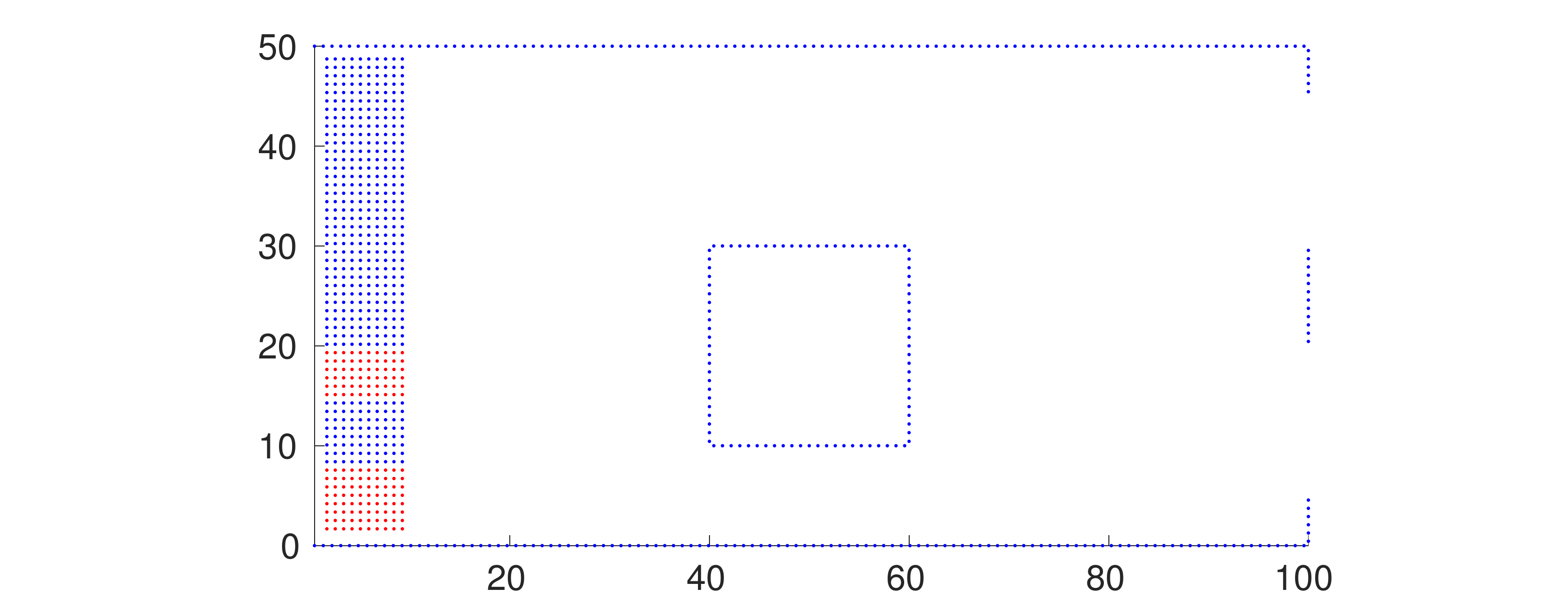}}
  \caption{Pedestrian groups at $t = 0$.}
\label{fig:modelinitial}
\end{figure}

For the following investigation on the  influence of parameters on the solution, we always consider
a situation as shown in Figure \ref{fig:modelinitial} with two groups. 
Alltogether, we use $N=500$ pedestrians initially located in $[0,10]\times [0,50]$ and $\rho= 1 $.
Group 1 is the larger one and  group 2 
the smaller one. In all cases group 1 consists of single pedestrians. Both groups are  split into two parts.

\subsubsection{Comparison between single and multi group pedestrian flow models with weak and strong reciprocal interaction}
For the numerical simulation  we use the following parameters.
The repulsive interaction is given by   $ l_r = 2 $ and $C_r = 100, 200$ respectively.
The attractive interaction uses   $ l_a = 4 $ and $C_a$ is chosen as  $C_a =0 $,  $C_a = 10$, $C_a = 50 $, $  C_a = 70 $.
Single pedestrians are modelled by $C_a=0$.
 
 
\begin{figure}
\subfloat
  {\includegraphics[width=.32\linewidth]{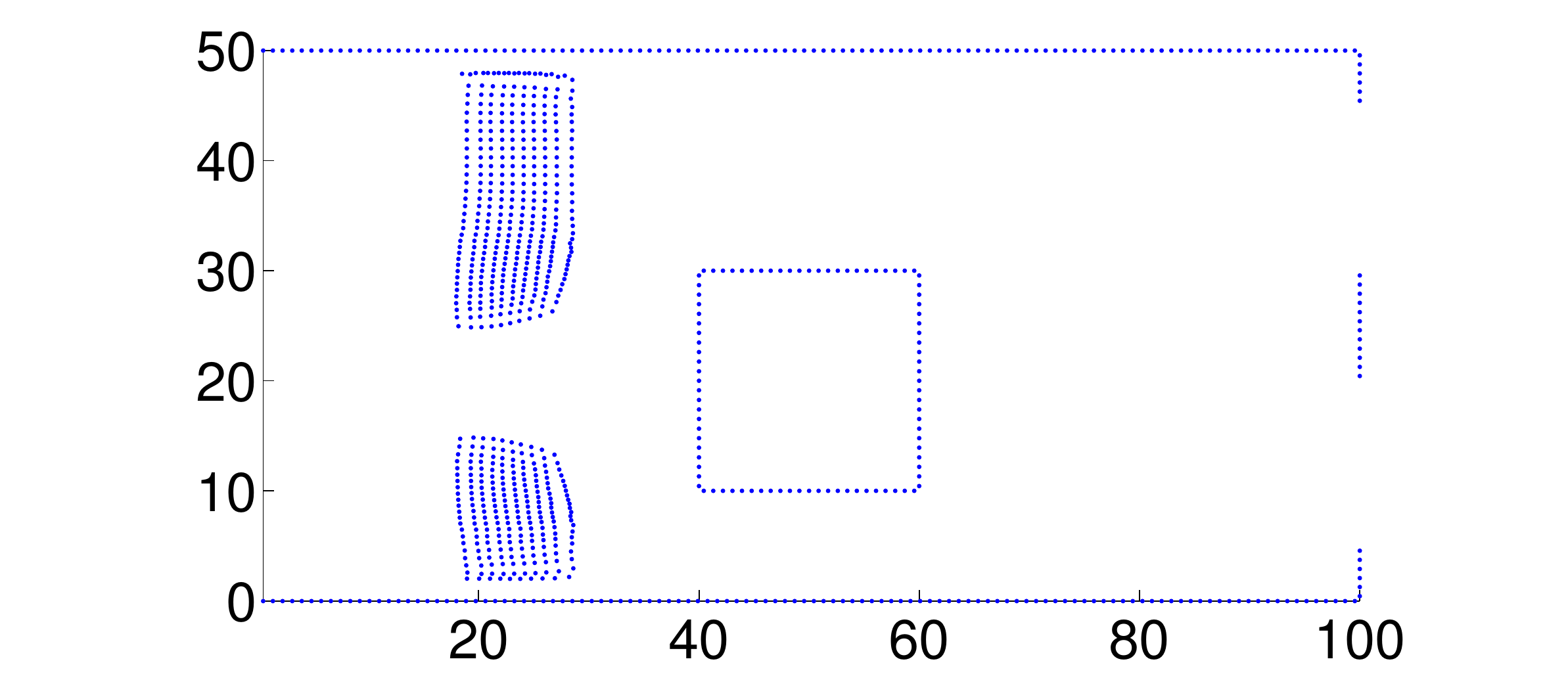}}
  \subfloat
  {\includegraphics[width=.32\linewidth]{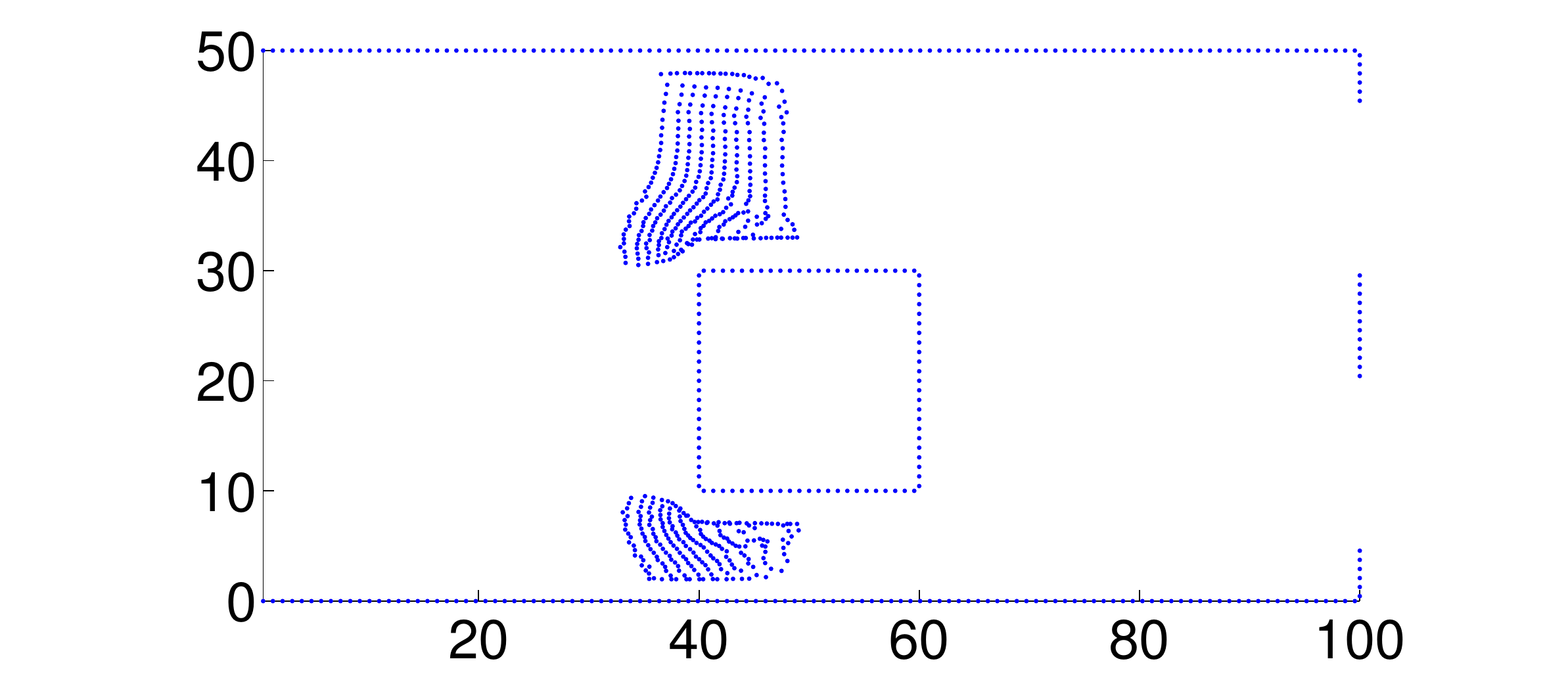}}
\subfloat
  {
  \includegraphics[width=.32\linewidth]{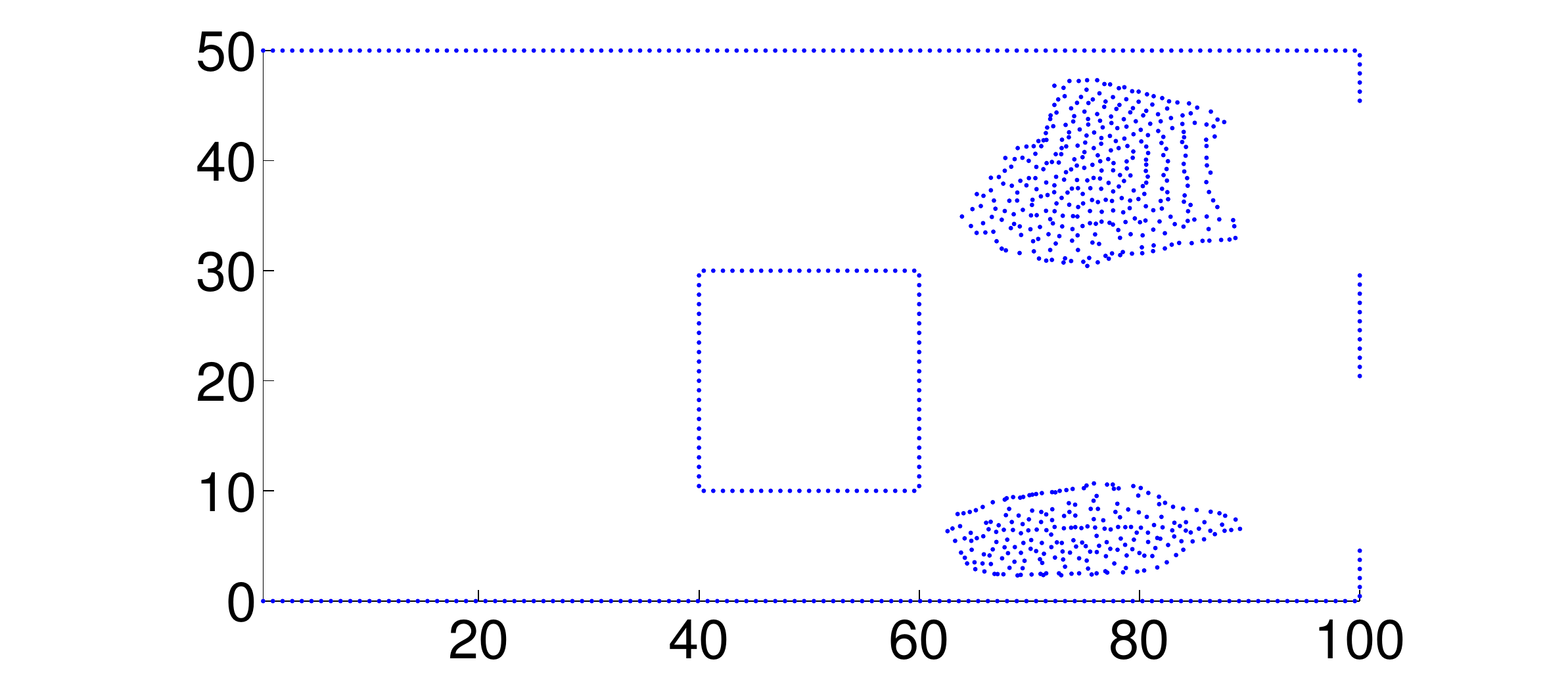}}\\
  \subfloat
    {\includegraphics[width=.32\linewidth]{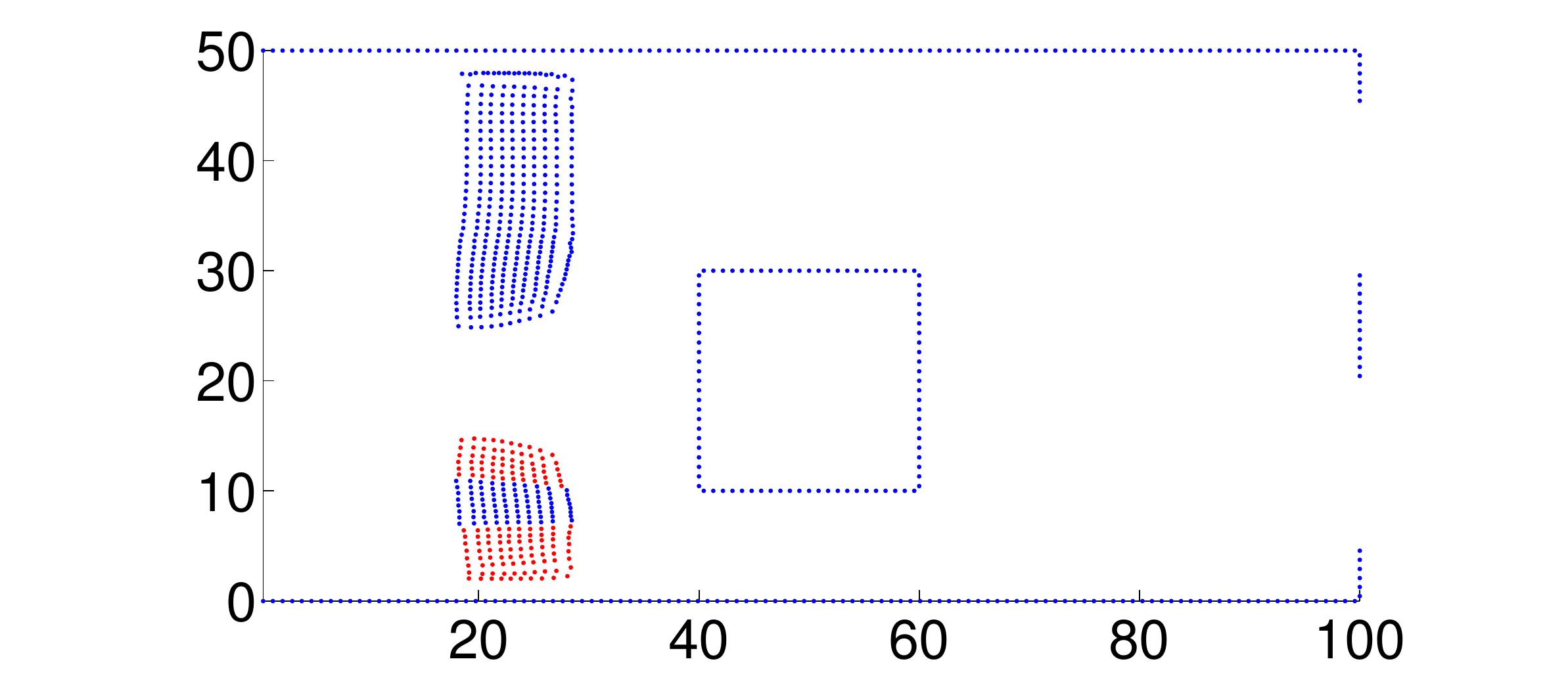}}
  \subfloat{
  \includegraphics[width=.32\linewidth]{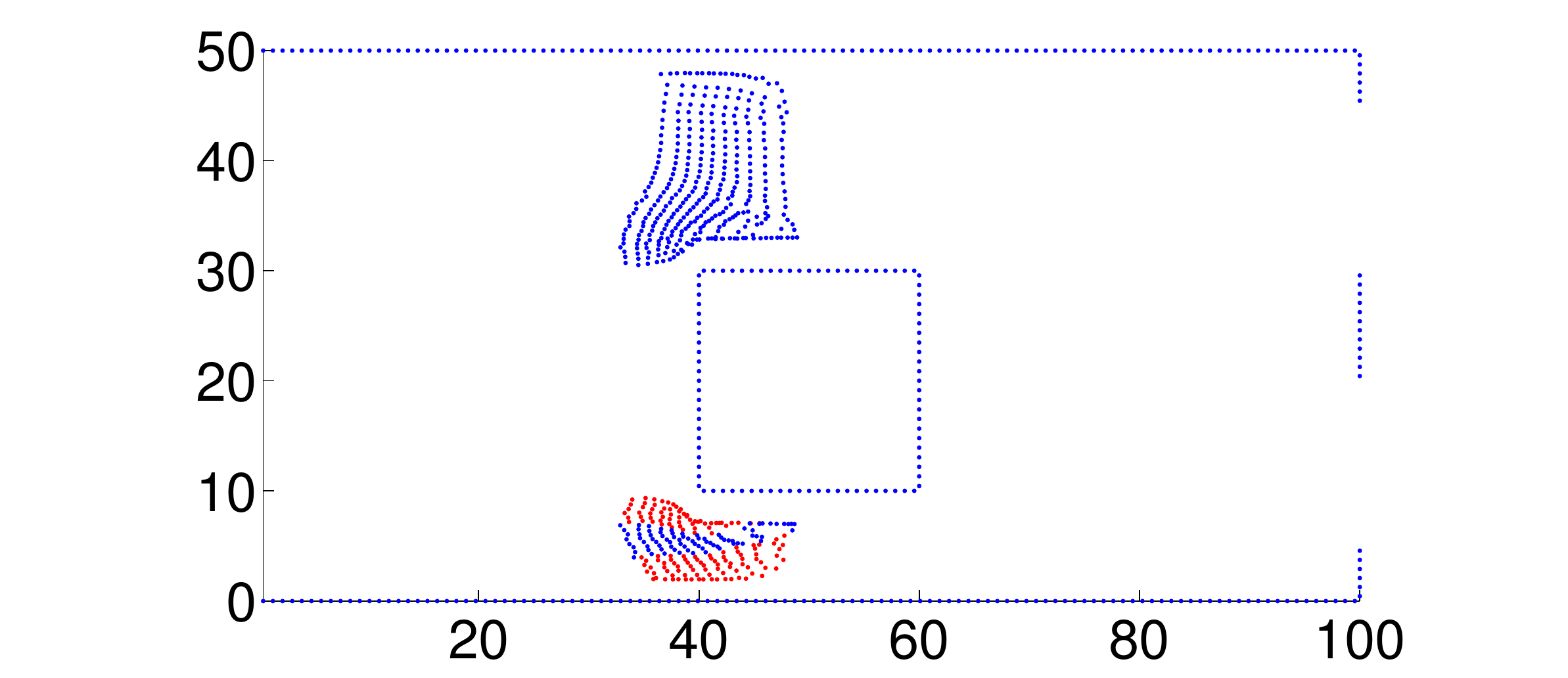}}
  \subfloat{
  \includegraphics[width=.32\linewidth]{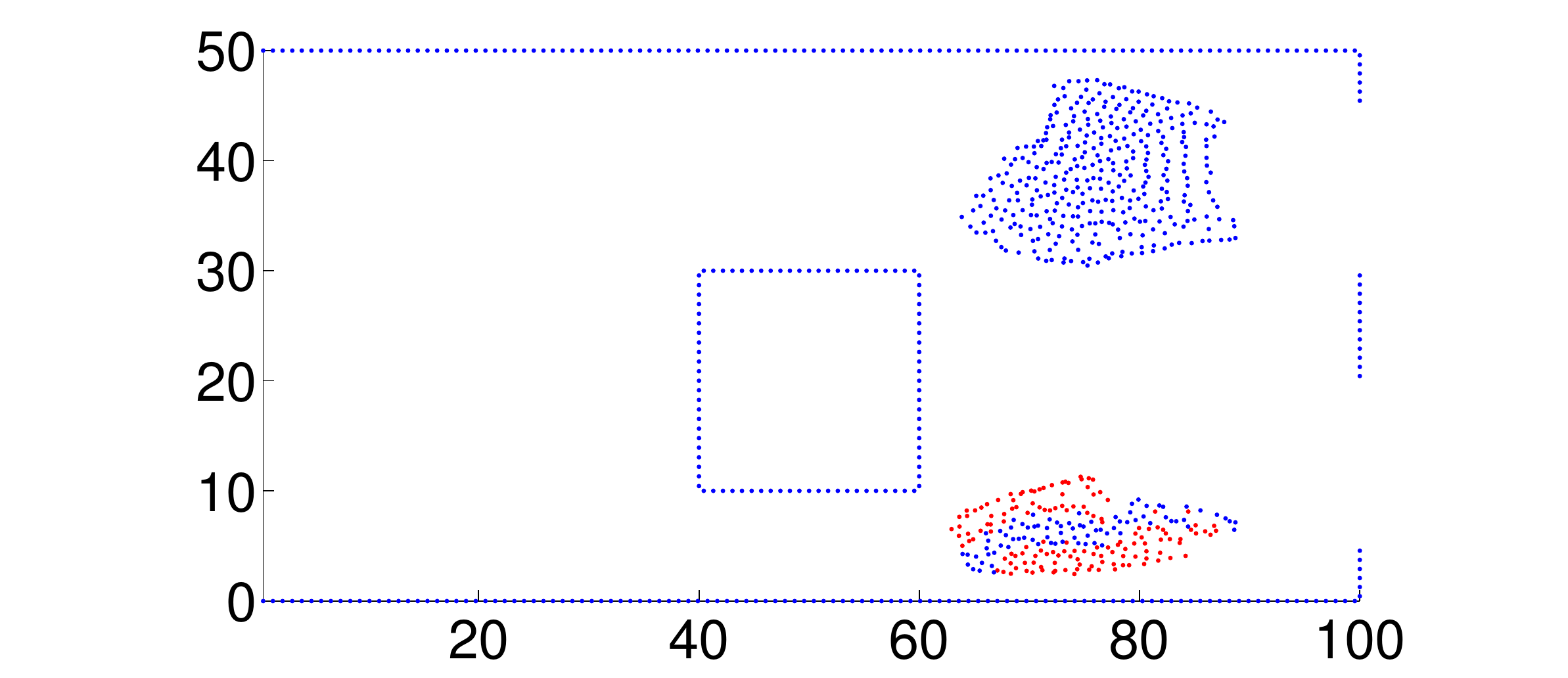}}\\
   \subfloat
      {\includegraphics[width=.32\linewidth]{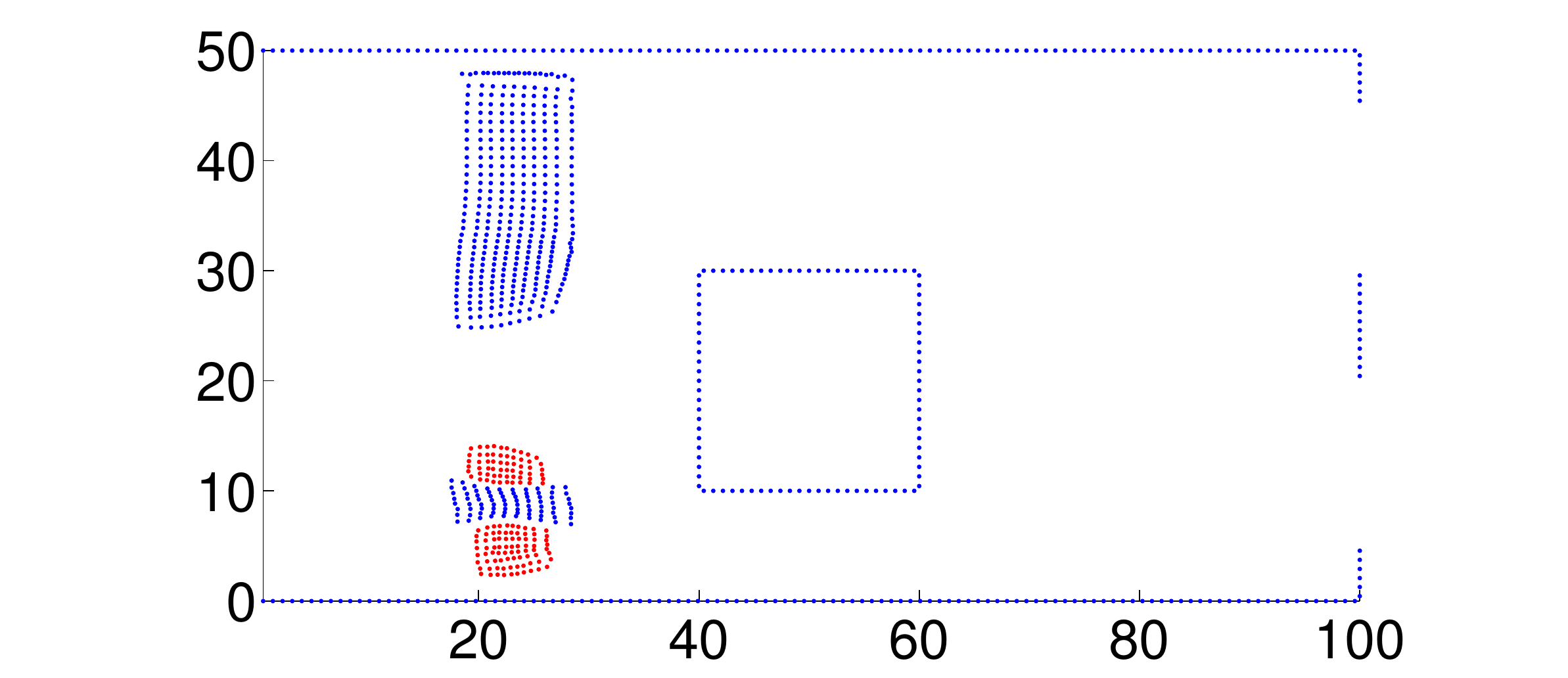}}
    \subfloat{
    \includegraphics[width=.32\linewidth]{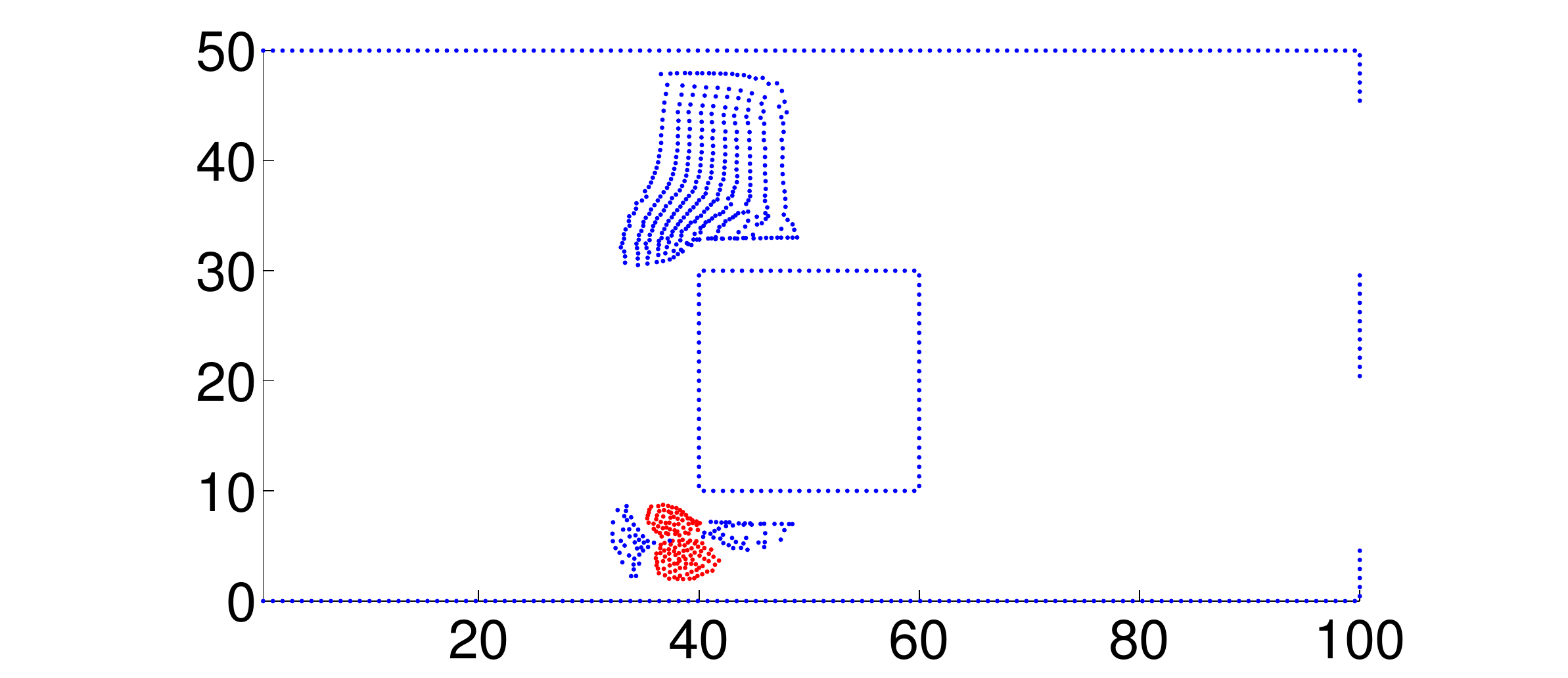}}
    \subfloat{
    \includegraphics[width=.32\linewidth]{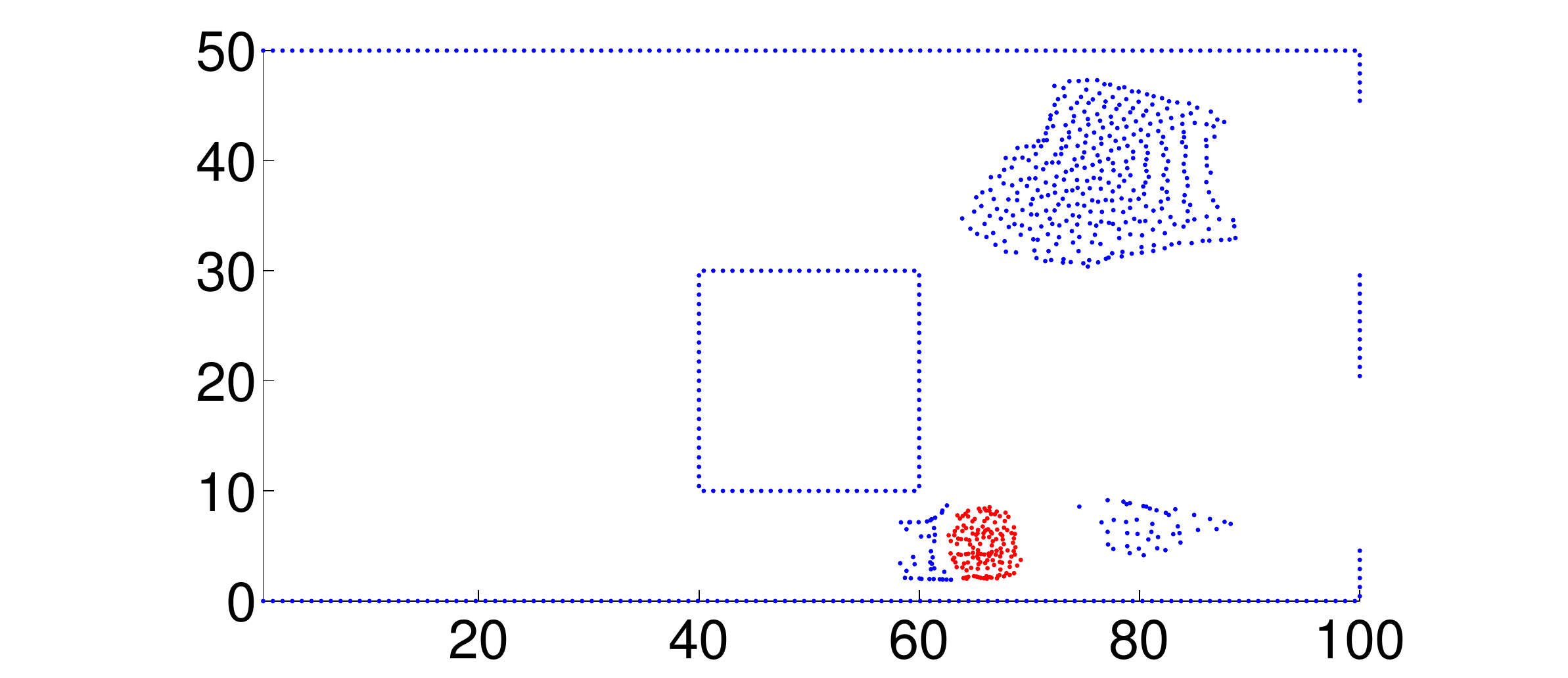}}\\
  \caption{Distribution of grid particles in single (first row) and multi-group  hydrodynamic models 
with $C_a =10$ (second row)  and $C_a =70$ (third row) for  different times $t = 10, 20,40$ (from left to right)}
\label{fig:2a}
\end{figure}

\begin{figure}[htbp]
\centering
\includegraphics[scale=0.3]{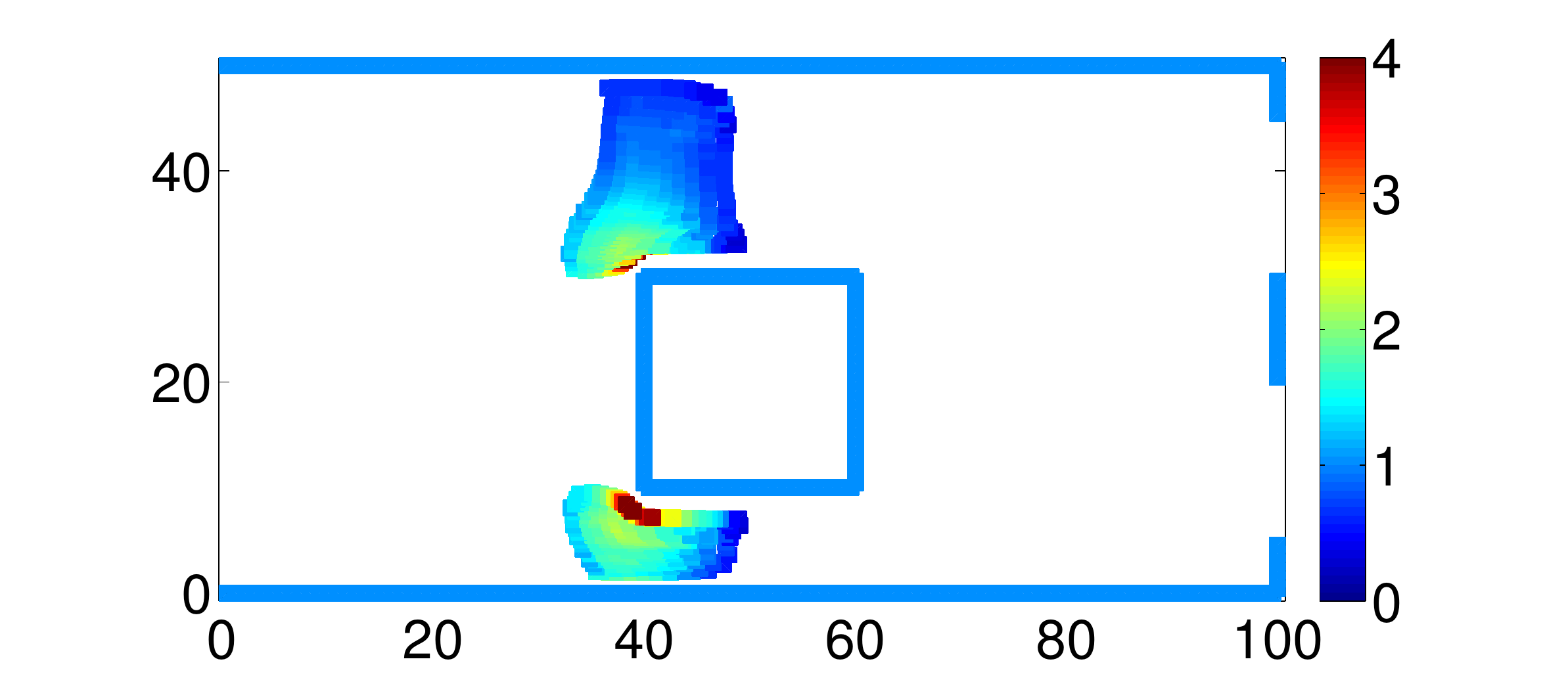}\\
\includegraphics[scale=0.3]{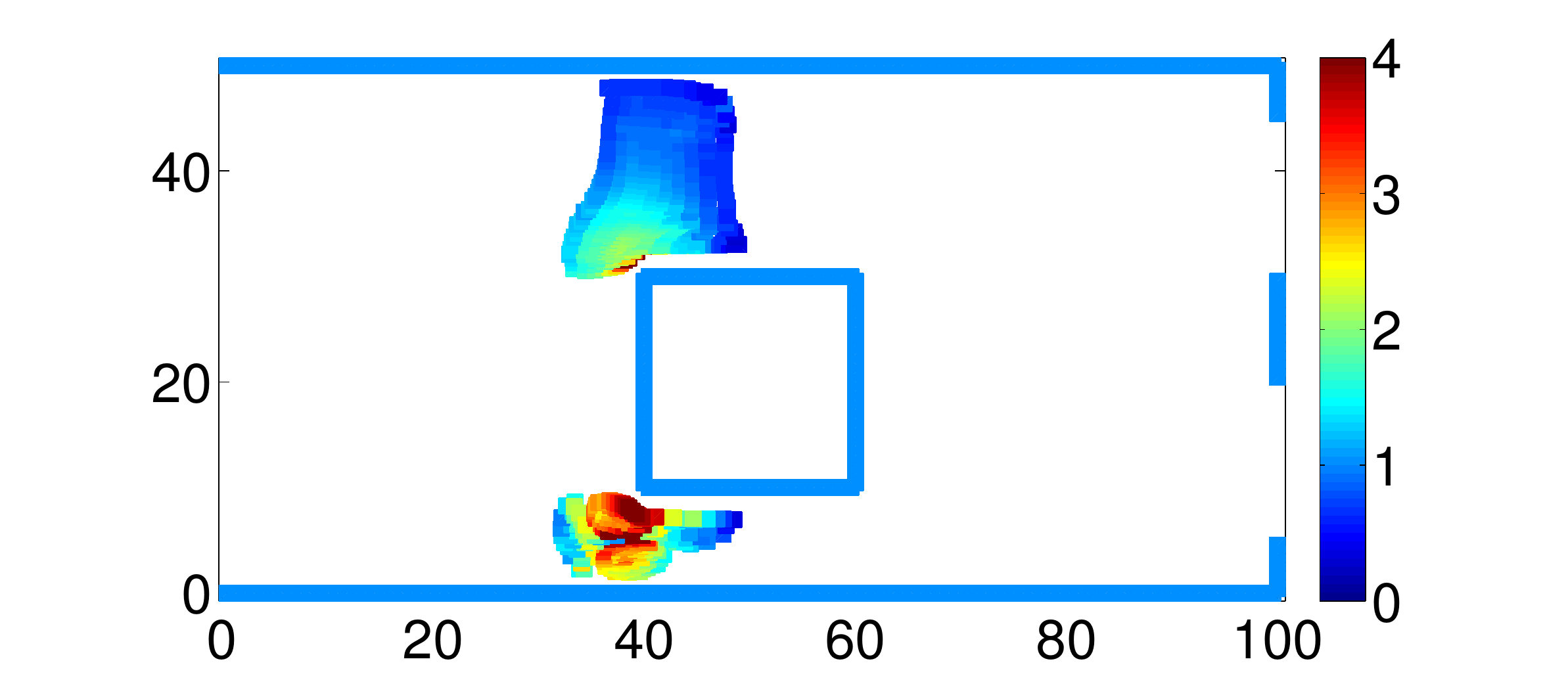}
\captionsetup{justification=centering}
\caption{Density of pedestrians for  single (top) and multi-group (bottom) hydrodynamic model at  times $t = 20$. }
\label{fig:2b}
\end{figure}

 Figures \ref{fig:2a} shows the time evolution of the grid particles in single and multi-group hydrodynamic models for time t = 10, t = 20 and  t = 40. Group 1 is modelled as single pedestrians with $C_a=0$, whereas for group 2 we use  $C_a = 0$ (all pedestrians are single), $C_a = 10$ and $C_a = 70$ respectively. One observes that single  pedestrians are faster than the multi-group pedestrians for stronger attractive interactions. In the multi-group model, grouped pedestrians with stronger attractive interactions walk slower compared to individual pedestrians or pedestrians with smaller attractive interaction. Some individual pedestrians become slower since the  grouped pedestrians  play the role of obstacles  for them.
Figure \ref{fig:2b} shows the corresponding density plots for the time $t=20$ for single and multi-group case
with $C_a = 70$.

\begin{figure}[htbp]
\centering
\subfloat[$C_r=200$]{\includegraphics[scale=0.3]{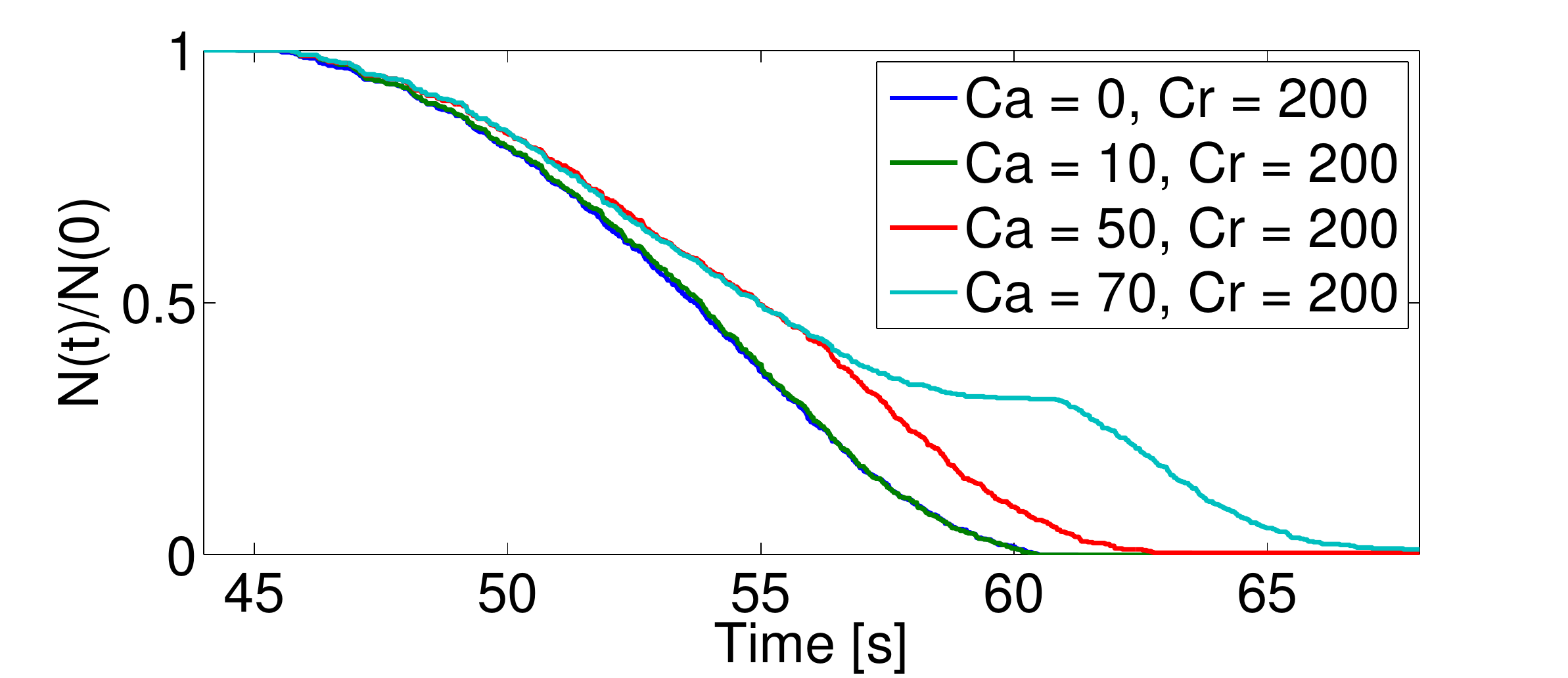}}\\
\subfloat[$C_r=100$]{\includegraphics[scale=0.3]{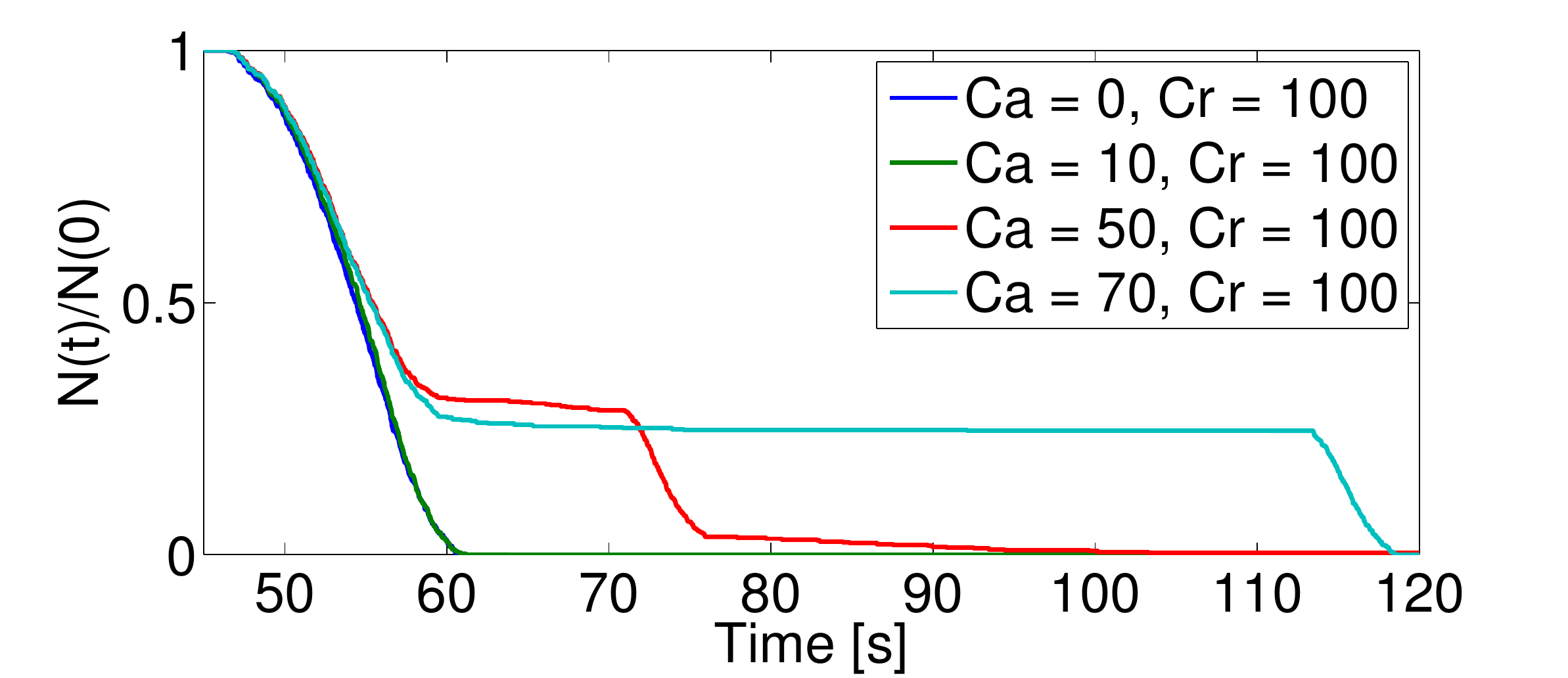}}
\captionsetup{justification=centering}
\caption{Ratio of initial and actual grid particles  with respect to time in single and multi-group hydrodynamic model with $C_a = 0,10,50,70$ for pedestrians in group 2. Figure (a)  shows the case $C_r =200$, Figure (b)  the case $C_r=100$.} 
\label{fig:3}
\end{figure}

 Figure \ref{fig:3} shows the percentage of grid particles being  in the computational domain  for single and multi-group hydrodynamic models with different interaction parameters with respect to time. One observes again, that the   evacuation time is larger  in the case of grouped pedestrians.  
  Moreover,  choosing the attraction coefficient in the above range one obtains  a monotonic behaviour: the evacuation times increase with increasing attraction.
  This corresponds  to the discussion in Remark \ref{discussion} and subsection \ref{exp}.
   Comparing the  Figures \ref{fig:3} (a) and \ref{fig:3} (b) one observes a similar trend for both cases, however, for smaller repulsive interaction, the pedestrians walking in group 2 become much slower leading to the plateau observed in Figure \ref{fig:3} (b).

 \subsubsection{Comparison between models with  weak and strong center of mass attraction} 
\begin{figure}[htbp]
  	\centering
  	\includegraphics[scale=0.3]{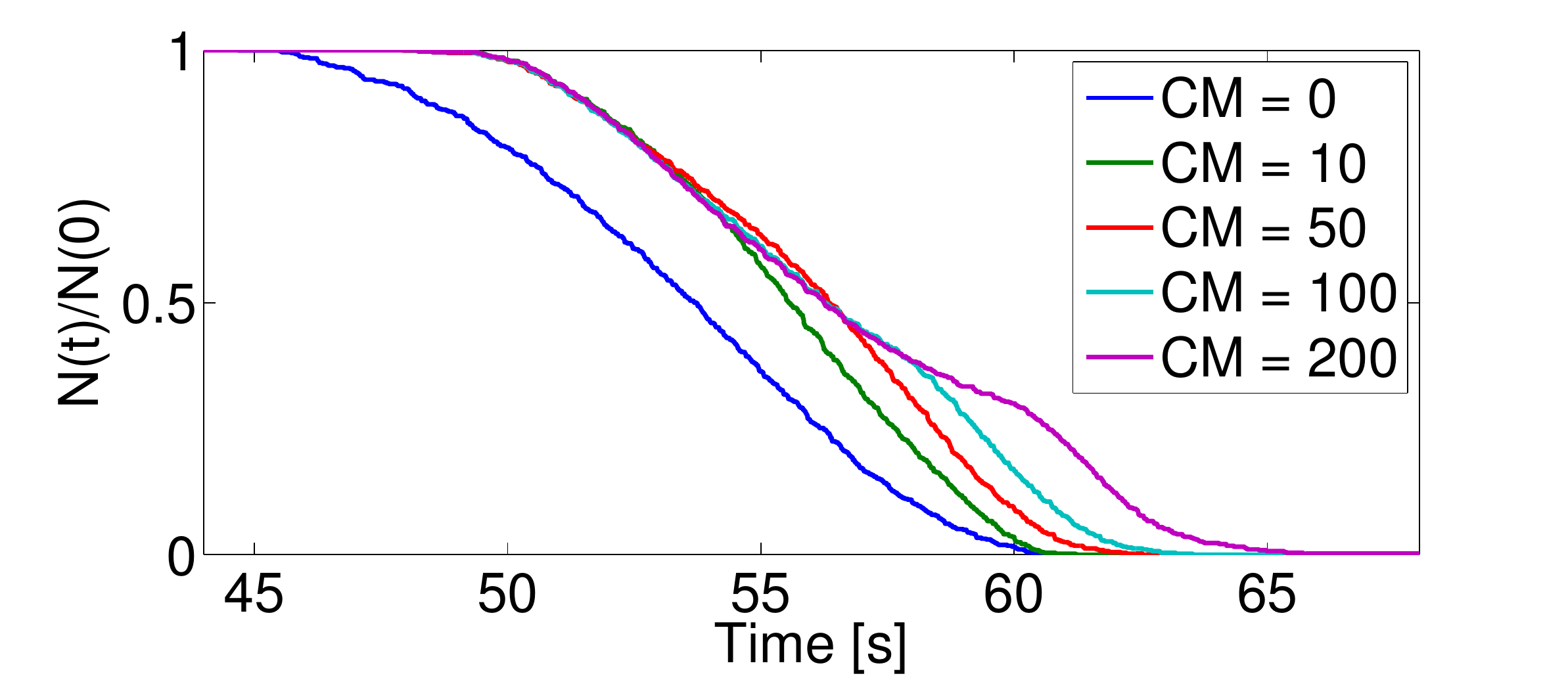}
  	\captionsetup{justification=centering}
  	\caption{Ratio of initial and actual grid particles in the computational domain with respect to time for multi-group hydrodynamic model with  center of mass attraction.}
  	\label{fig:9}
  \end{figure}

 For the numerical simulation, we use  a quadratic repulsive interaction potential $U(x) = C  (2 R -\vert x \vert)^2$ 
 with  parameters  $C = 1000$ and $R=0.4$. 
 Figure \ref{fig:9} shows the number of grid particles in the computational domain  for the case with center of mass interaction.
 We use as potential $U_{CM}(x) = -C_{CM}\vert x \vert^2$ yielding a linear force towards the center of mass.
 In Figure \ref{fig:9} we display 5 results;   the case without center of mass attraction and  the cases with
 $ C_{CM} =  10,50$, $100$ and $200$.
 Obviously, the center of mass attraction has a similar influence as the  reciprocal interaction
 in the above subsection. Choosing the parameters in the above range, one obtains  again a  monotone behaviour 
 for the evacuation times.

\subsubsection{Comparison  between models with different relaxation times}
In this subsection we compare the evacuations time of grouped and individual pedestrians for three different relaxation times $T = 0.1, 0.01$ and $0.001$. In all three cases we use the parameters: $C_a = 50$,  $ l_a = 5 $,  $ l_r = 2 $ and  two different values of $C_r$ that is $C_r = 100$ and $C_r = 200$. 

\begin{figure}
 \subfloat
   {\includegraphics[width=.32\linewidth]{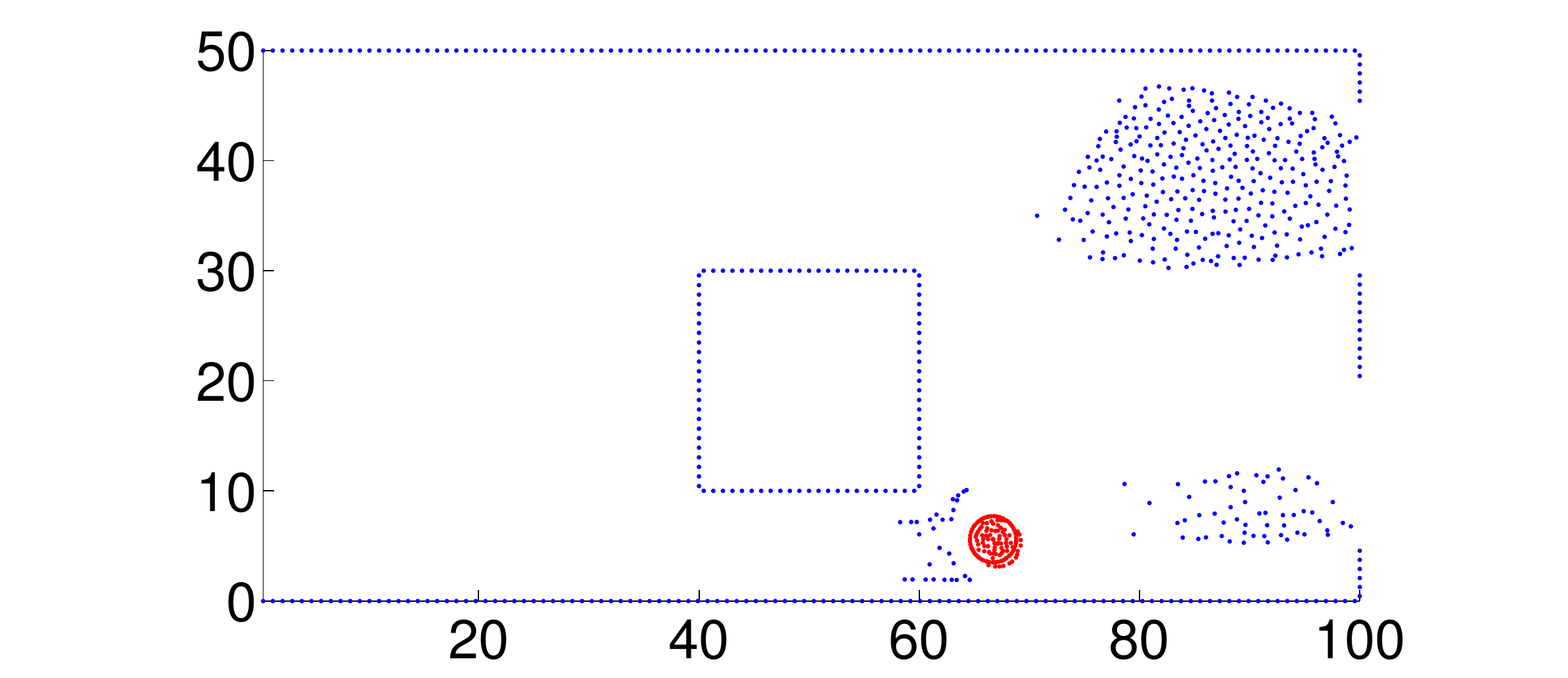}}
   \subfloat
   {\includegraphics[width=.32\linewidth]{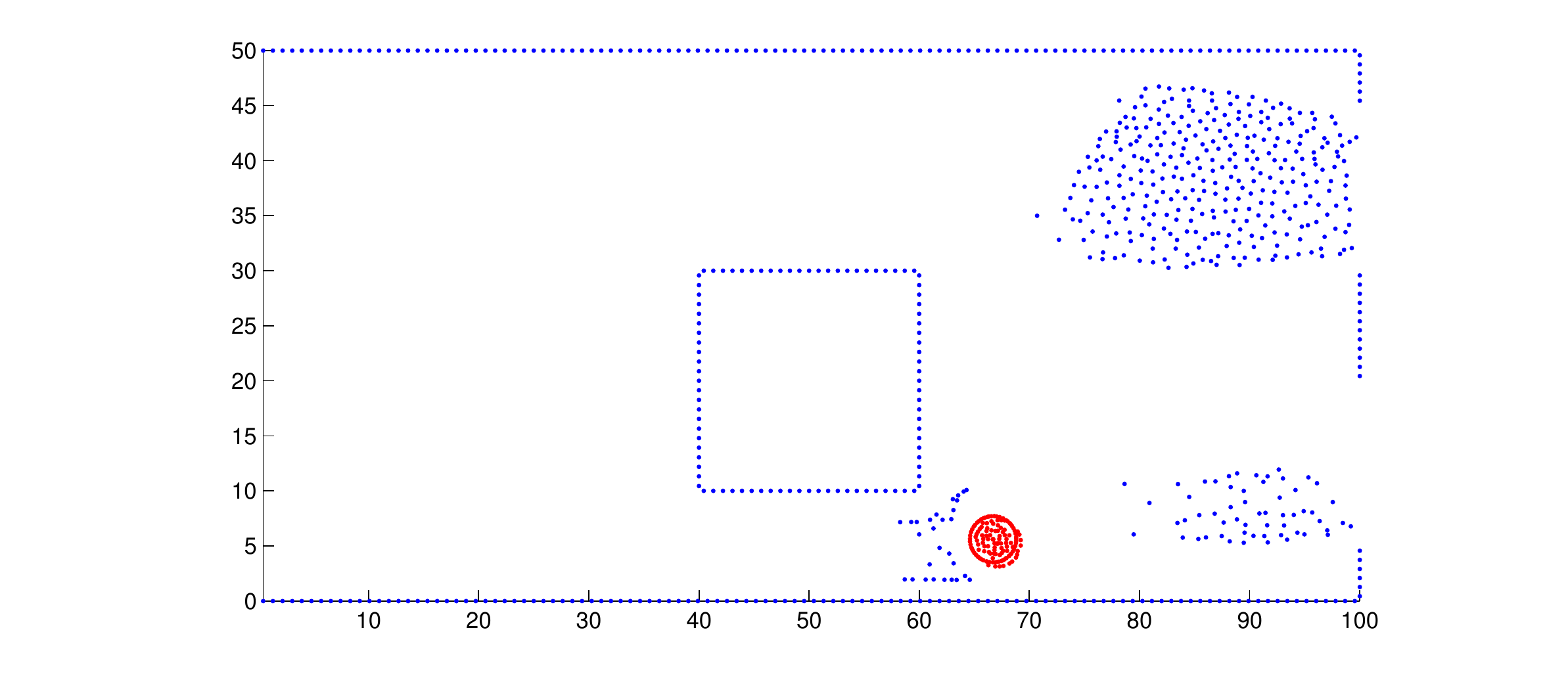}}
 \subfloat
   {
   \includegraphics[width=.32\linewidth]{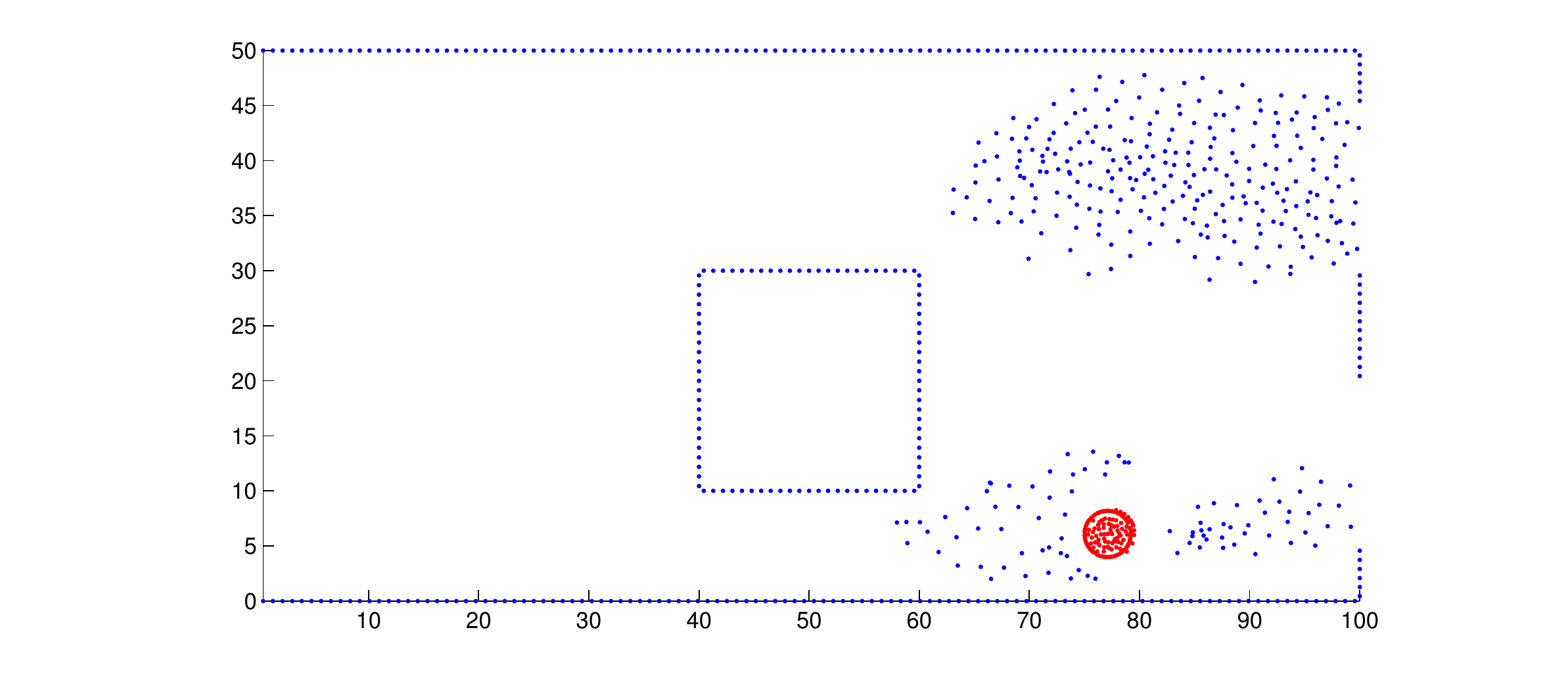}}\\
   \subfloat
     {\includegraphics[width=.32\linewidth]{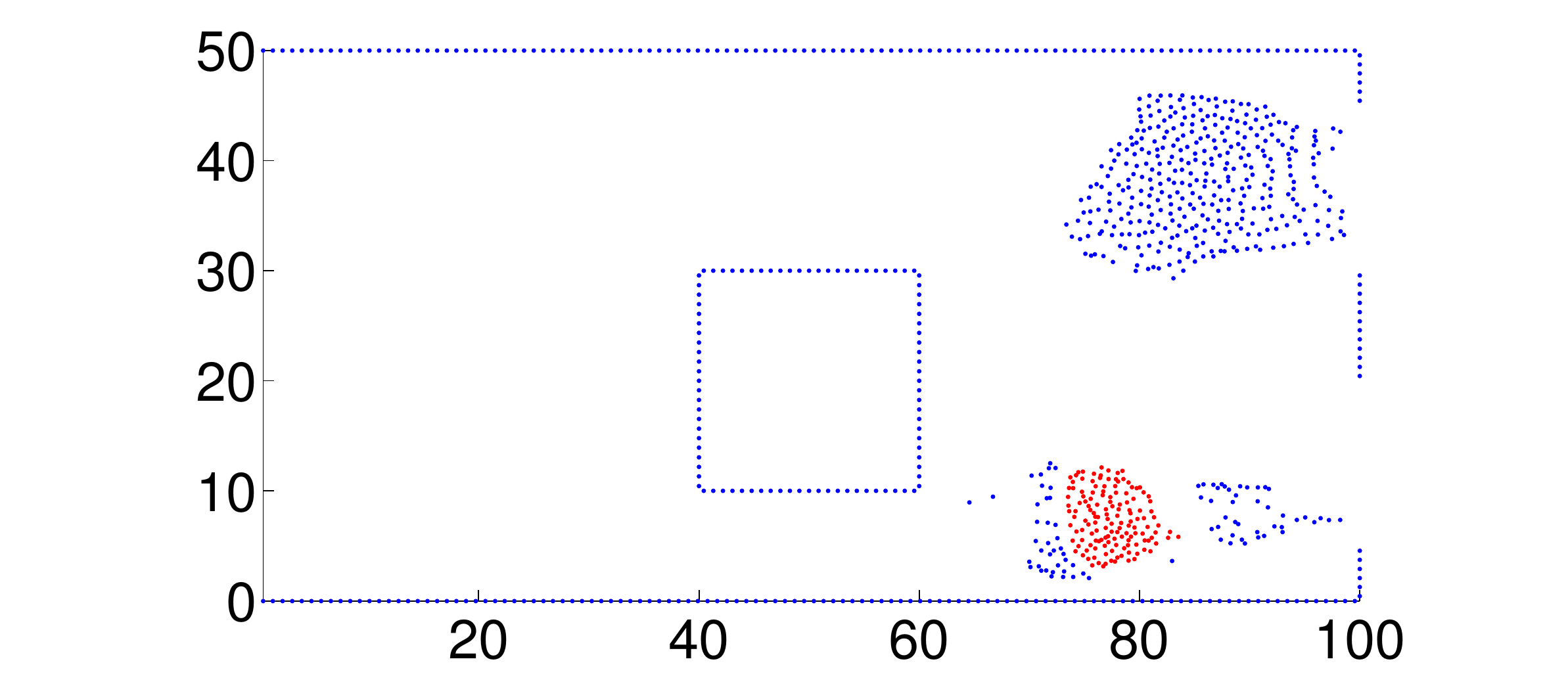}}
   \subfloat{
   \includegraphics[width=.32\linewidth]{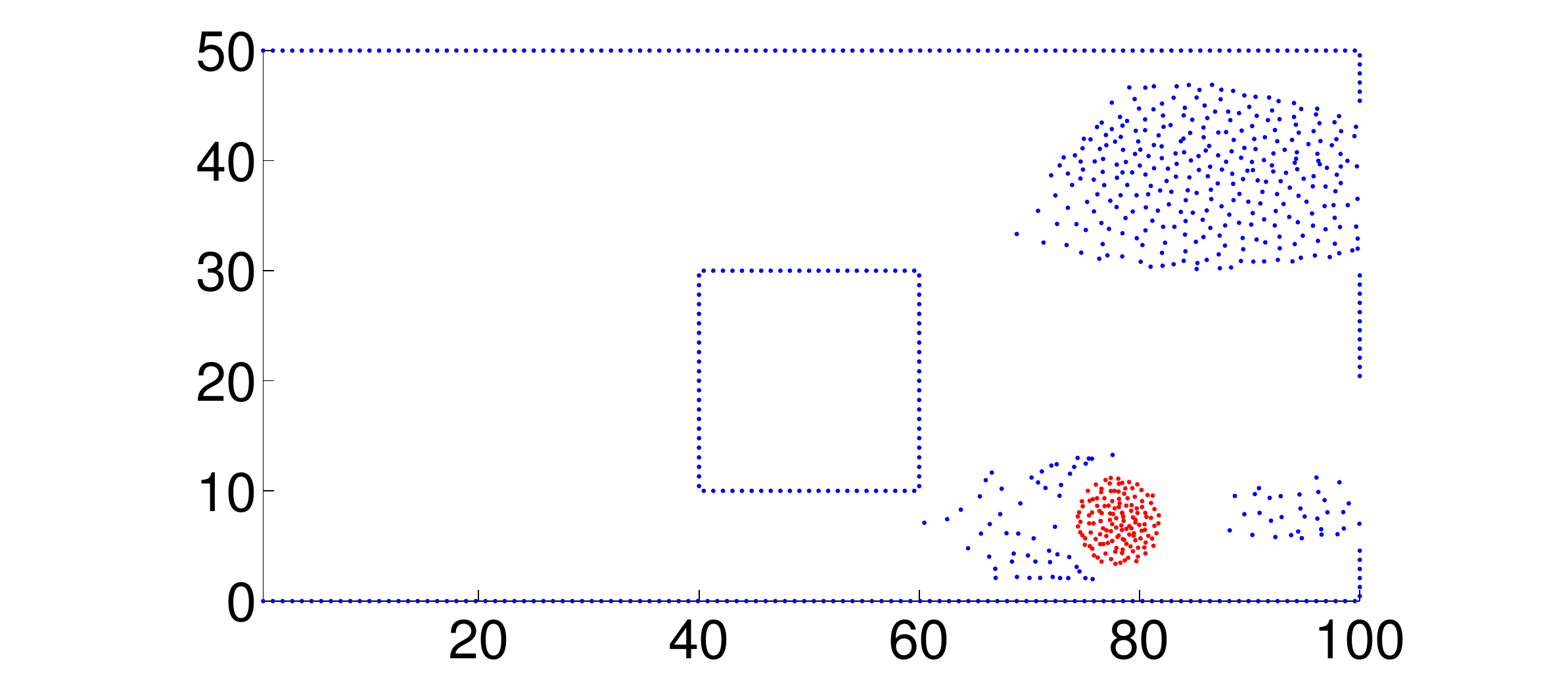}}
   \subfloat{
   \includegraphics[width=.32\linewidth]{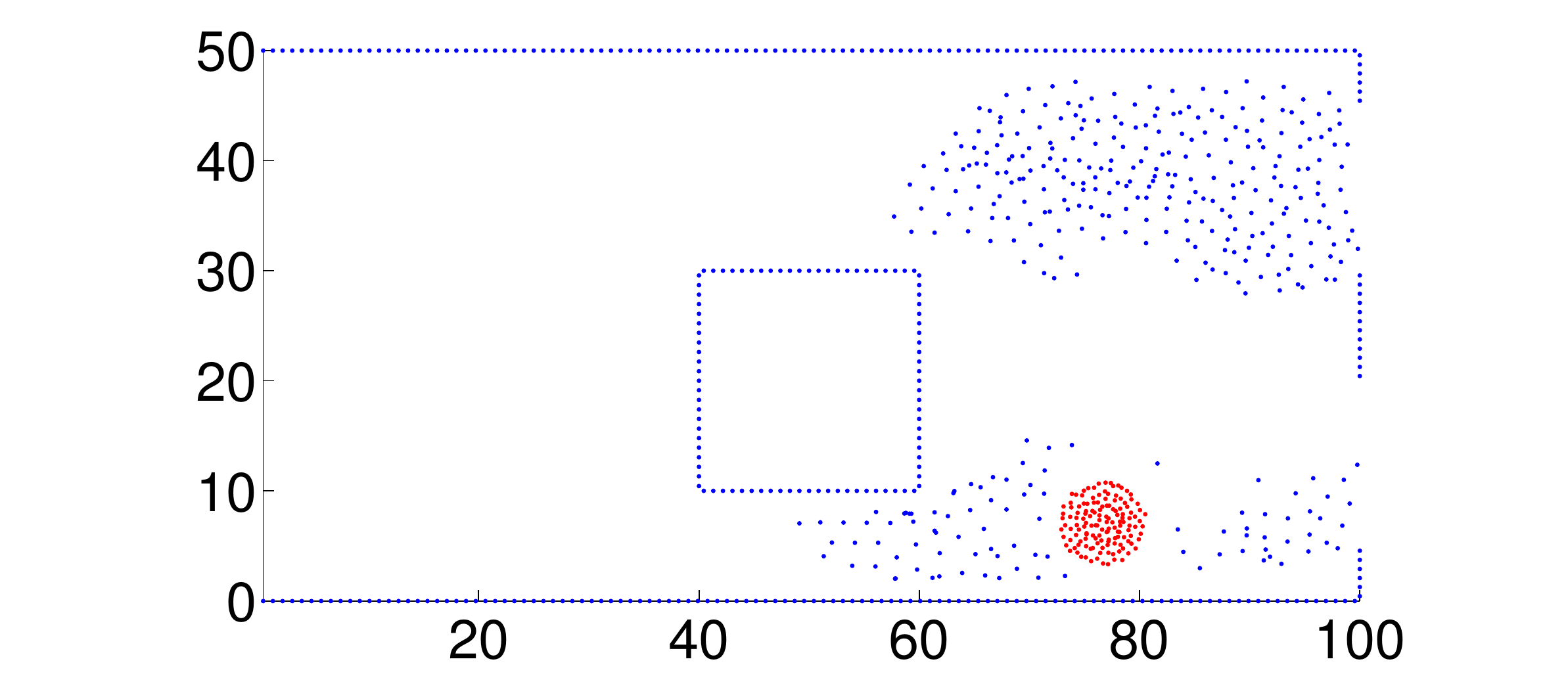}}\\
  \caption{Distribution of grid particles for  $C_r = 100$ (first row) and $C_r = 200$ (second row) for  hydrodynamic model
 at  time $t = 45$ for different values of $T = 0.001, 0.01, 0.1$ (from left to right)}
 \label{fig:compareTa}
 \end{figure}

\begin{figure}[htbp]
 \centering
 \subfloat[$C_r=100$]{\includegraphics[scale=0.32]{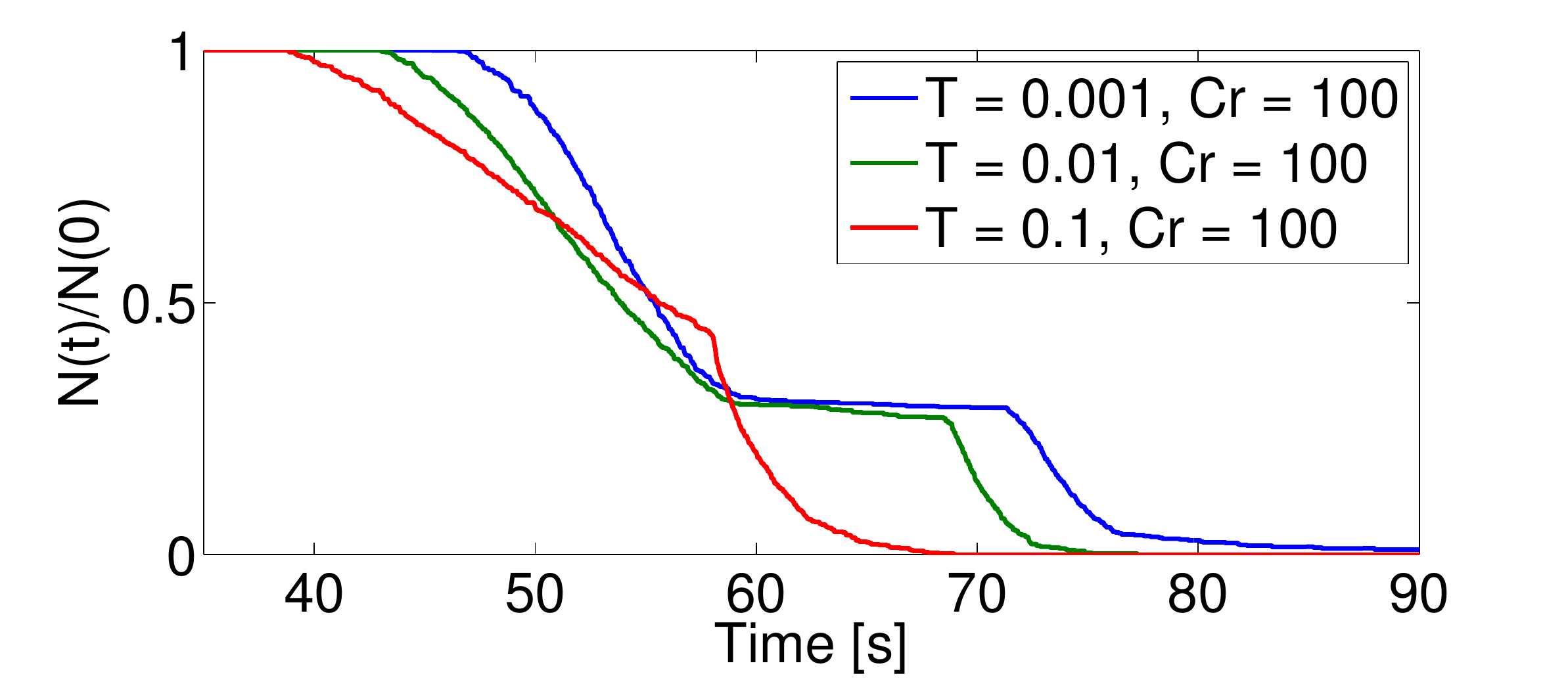} }\\
 \subfloat[$C_r=200$]{\includegraphics[scale=0.32]{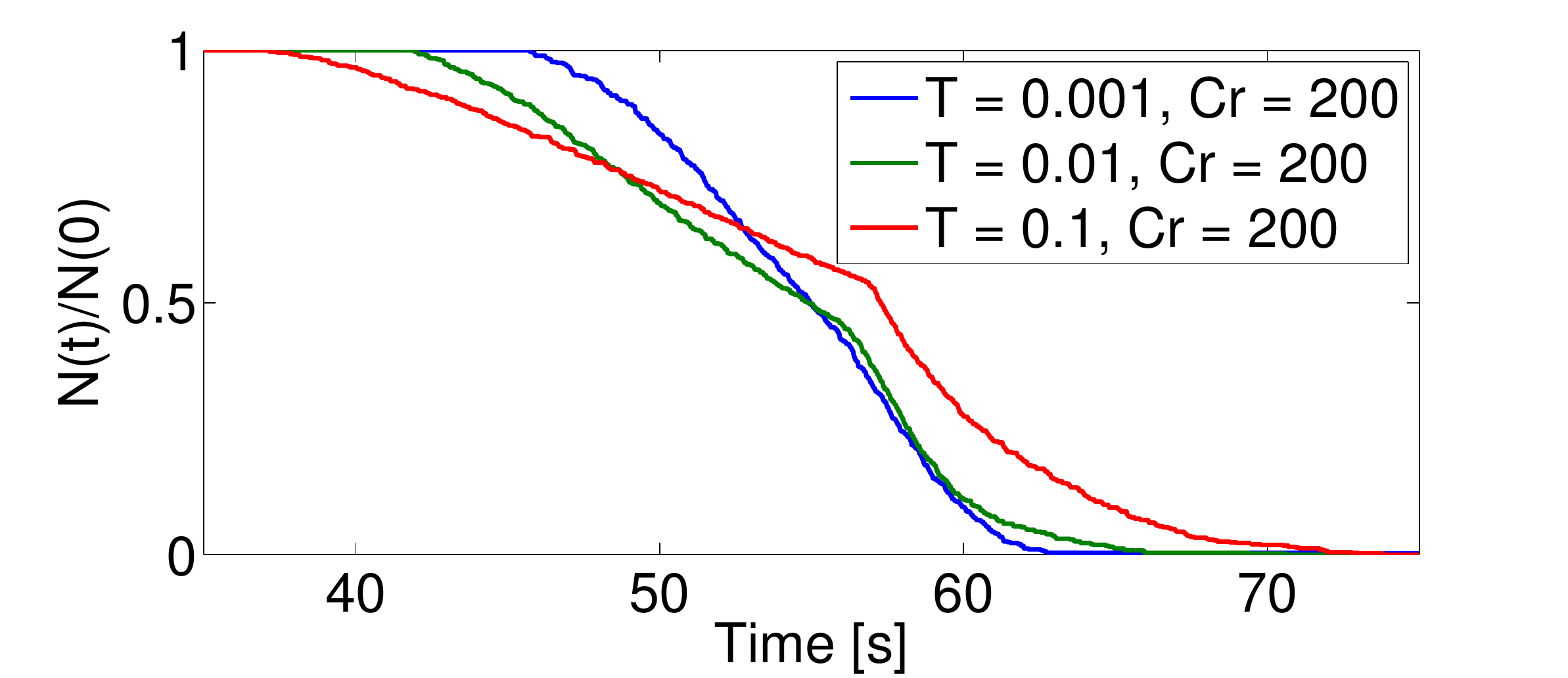}}
  \captionsetup{justification=centering}
 \caption{Ratio of initial and actual grid particles  in the computational domain with respect to time for multi-group hydrodynamic model for $T=0.001, T=0.01$ and $T=0.1$ with $C_r = 100$ and $C_r = 200$}
 \label{fig:compareTb}
 \end{figure}

In Figure \ref{fig:compareTa} we have plotted the position of the grid particles for $C_r = 100$ in the first row and $C_r = 200$ in the second row at time $ t = 45$ 
for $T = 0.001, 0.01 $ and $T = 0.1$. 
Figure \ref{fig:compareTb} shows the time evolution of the number of grid particles in the computational domain for  hydrodynamic equations for $T=0.001$, $T=0.01$ and  $T=0.1$.
We observe that increasing $T$ disperses the pedestrians and leads to a larger velocity of the pedestrians
until the first pedestrians arrive at the exits, 
compare the discussion in Remark \ref{discussion}. However, the behaviour of the exit times in figure \ref{fig:compareTb} (b) is not any more monotone. In this case  the behaviour is dominated by other effects like for example the fact that  the  grouped pedestrians  play the role of obstacles  for the single pedestrians following them.
  
\subsection{Experimental data}
\label{exp}
The effect of grouping of pedestrians on evacuation processes, and in particular on the evacuation time, has been considered in a series of recent publications from an experimental point of view, as well as with the help of numerical experiments. We refer to 
\cite{KS17, Koe14} for experimental data in simplified situations. \cite{group2} discusses the walking speed for groups of different sizes and
\cite{Chu14} and \cite{group1} use   an agent based model to investigate social groups in pedestrian flow.
Larger groups are, for example, considered in \cite{XZCH14}.
In \cite{K14}, besides showing  results on evacuation times,  a comparison and a critical discussion  of previous approaches is given.

The experimental results in \cite{Koe14} consider only small groups. Nevertheless, they show   an increase in evacuation time for an increase in group size. Similarly, in \cite{KS17} the authors obtain  for groups with cooperative behaviour  a longer evacuation time than for individuals.
Such a  trend can also be seen in \cite{Chu14}  and \cite{K14}.
In particular, in the last two papers, larger groups are shown  to have up to  $50 \% $ longer evacuation times than individuals. This is in accordance with our results giving  ecvacuation times
for grouped motion,
which are up to $50 \%$ longer, compare the results in Figure  \ref{fig:3}  for not too large values of $C_a$.

\section{Concluding Remarks}
We have presented a multi-group microscopic model combining a  social force model and an optimal path computation for  pedestrians flows.
Hydrodynamic and scalar models are derived from the microscopic model.
 A meshfree particle method to solve the governing equations is presented and used for the  computation of several numerical example analysing  single-, and multi-group hydrodynamic  models for different parameters such as interaction coefficients and relaxation time.
 The dependence of the solutions on these parameters is investigated and discussed.
 As a general result we observe for our examples that increasing   the attraction between the group members  increases evacuation time. Increasing the relaxation time does not necessarily lead to a decrease in evacuation times. The behaviour in this case may be dominated by other effects.  
 
  Future research topics are in particular the  consideration of more complex situations and a more detailed  identification procedure for  the 
  parameters in  the above models.

\section*{Acknowledgment}

This work is supported by the German research foundation, DFG grant KL 1105/27-1, by RTG GrK 1932 “Stochastic Models for Innovations in the Engineering Sciences”, project area P1 and  by the
DAAD PhD program MIC "Mathematics in Industry and Commerce".


\begin{thebibliography}{30}

\bibitem{ama}
D. Amadoria and M. Di Francesco,
\textit{The one-dimensional Hughes model for pedestrian flow: Riemann-type solutions},
Acta Math. Sci. 32 (2012) 259-280.

\bibitem{bello}
N. Bellomo and C. Dogbe, 
\textit{On the modeling crowd dynamics from scaling to hyperbolic macroscopic models},
Math. Models Methods Appl. Sci. 18 (2008) 1317-1345.

\bibitem{bell}
N. Bellomo and C. Dogbe, 
\textit{On the modeling of traffic and crowds: A survey of models, speculations, and perspectives},
SIAM Rev. 53 (2011) 409-463.

 
\bibitem{BBK13} 
N. Bellomo, A. Bellouquid, D. Knopoff,
From the microscale to collective crowd dynamics,
SIAM J. Multiscale Model. Simul. 11 (2013) 943--963.

\bibitem{∫BKM14}
R. Borsche, A. Klar, S. K\"uhn, A. Meurer, Coupling traffic flow networks to pedestrian motion, Math. Methods Models Appl. Sci. 24, 2, 359-380, 2014

\bibitem{braun1977vlasov}
W~Braun and K~Hepp.
\newblock The {V}lasov dynamics and its fluctuations in the 1/n limit of
  interacting classical particles.
\newblock {\em Communications in mathematical physics}, 56(2):101--113, 1977.

\bibitem{B01}
C. Burstedde, K. Klauck, A. Schadschneider, J. Zittartz,
Simulation of pedestrian dynamics using a two-dimensional cellular automaton,
Physica A 295  507–525, 2001.


\bibitem{canizo2011well}
J. A Canizo, J. A Carrillo, and J.  Rosado.
\newblock A well-posedness theory in measures for some kinetic models of
  collective motion.
\newblock {\em Mathematical Models and Methods in Applied Sciences},
  21(03):515--539, 2011.




\bibitem{hyd}
J.A. Carrillo, M.R. D'Orsogna, and V. Panferov,
\textit{Double milling in self-propelled swarms from kinetic theory},
Kinetic and Related Models, vol 2, no 2, (2009), pp. 363-378.


\bibitem{roosting} 
J.A. Carrillo, A. Klar, S. Martin and S. Tiwari, 
\textit{Self-propelled interacting particle systems with roosting force}, 
Mathematical Models and Methods in Applied Sciences 20, Suppl., 1533-1552, 2010. 

\bibitem{group1}
L. Cheng, V. Reddy, C. Fookes, and P. K. D. V. Yarlagadda,
\textit{Impact of passenger group dynamics on an airport evacuation process using an agent-based model}, 
International Conference on Computational Science and Computational Intelligence, Las Vegas, Nevada, USA (2014).

\bibitem{Chu14}
M.L. Chu1, P. Parigi, K. Law, J.-C. Latombe,
Modeling social behaviors in an evacuation simulator,
Computer animation and virtual worlds, Comp. Anim. Virtual Worlds 2014; 25:375–384


\bibitem{Col1}
R. Colombo, M. Garavello, M. Lecureux-Mercier, 
A Class of Non-Local Models for Pedestrian Traffic, 
Math. Models Methods Appl. Sci. 22 (2012) 1150023. 

\bibitem{Col2}
R. Colombo, M. Garavello, M. Lecureux-Mercier,
Non-local crowd dynamics, 
Comptes Rendus Mathmatique 349 (2011) 769--772. 



\bibitem{deg}
P. Degond, C. Appert-Rolland, M. Moussaid, J. Pettre and G. Theraulaz,
\textit{A hierarchy of heuristic-based models of crowd dynamics},
J. Stat. Phys. 152 (2013) 1033-1068.

\bibitem{di}
M. Di Francesco, P.A. Markowich, J.F. Pietschmann and M.T. Wolfram,
\textit{On the Hughes model for pedestrian flow: The one-dimensional case},
J. Differential Equations 250 (2011) 1334-1362.

\bibitem{di2}
M. Di Francesco, S. Fagioli, M.D. Rosini, G. Russo,
Deterministic particle approximation of the {H}ughes model in one space dimension,
Kinetic and Related Models, 10, 1, (2017), 215–237




\bibitem{dobrushin1979vlasov}
R.L. Dobrushin.
\newblock Vlasov equations.
\newblock {\em Functional Analysis and Its Applications}, 13(2):115--123, 1979.





\bibitem{etikyala2014particle}
R~Etikyala, S~G{\"o}ttlich, A~Klar, and S~Tiwari.
\newblock Particle methods for pedestrian flow models: From microscopic to
  nonlocal continuum models.
\newblock {\em Mathematical Models and Methods in Applied Sciences}, 20(12), 
  2503--2523, 2014.


\bibitem{raghu3} 
R. Etikyala, S. G\"ottlich, A. Klar and S. Tiwari, 
\textit{A microscopic model for pedestrian flow:  comparisons with experimental results of pedestrian flow in corridors and T-junctions}, 
submitted to Neural, parallel and Scientific Computations [CNLS-2013].
 
\bibitem{greengard}
L. Greengard and V. Rokhlin,
\textit{A Fast Algorithm for Particle Simulations},
Journal of Computational Physics 135, 280-292 (1997). 

\bibitem{H66}
 E.T., Hall, The Hidden Dimension. Anchor Books. ISBN 0-385-08476-5, 1966.
 
 
\bibitem{helbing}
D. Helbing,
\textit{A fluid dynamic model for the movement of pedestrians},
Complex Syst. 6 (1992) 391-415.
 
\bibitem{helbing1} 
D. Helbing, 
\textit{Traffic and related self-driven many-particle systems}, 
Rev. Modern Phys, 73(4) (2001), pp.1067-1141.

\bibitem{he&Mo}
D. Helbing and P. Molnar,
\textit{Social force model for pedestrian dynamics},
Phys. Rev. E, 51 (1995), pp. 4282-4286.
 
\bibitem{helbing2} 
D. Helbing, I.J. Farkas, P. Molnar, and T. Vicsek,
\textit{Simulation of pedestrian crowds in normal and evacuation situations, in: M. Schreckenberg, S.D. Sharma(Eds.)},
Pedestrian and Evacuation Dynamics, Springer-Verlag, Berlin, 2002, pp. 21-58. 

\bibitem{HJ10} 
D. Helbing, A. Johansson, Pedestrian, Crowd and Evacuation Dynamics. Encyclopedia of Complexity and Systems Science 16, (210), 6476-6495.


\bibitem{hughes}
R.L. Hughes,
\textit{A continuum theory for the flow of pedestrians},
Transp. Res. Part B: Methodological 36 (6) (2002), pp. 507-535.

\bibitem{hughes1}
R.L. Hughes,
\textit{The flow of human crowds},
Ann. Rev. Fluid Mech. 35 (2003) 169-182.

\bibitem{KS17}
C. v.  Kr\"uchten, A. Schadschneider,
Empirical study on social groups in pedestrian evacuation
dynamics, Physica A 475 (2017) 129–141


\bibitem{KGT09}
  D.P. Kennedy, J. Gl\"ascher, J.M. Tyszka, R. Adolphs, Personal space regulation by the human amygdala. Nat Neurosci. 12, 1226–1227, 2009.
  
\bibitem{eikonal} 
A. Klar, S. Tiwari, and E. Raghavender, 
\textit{Mesh Free method for Numerical Solution of The Eikonal Equation, Proceedings of International workshop on PDE Modelling and Computation},
Advances in PDE Modelling and Computation, Ane Books Pvt. Ltd., 2013. 


\bibitem{KT14}
A. Klar, S. Tiwari,  A multi-scale  meshfree particle method for macroscopic mean field  interacting particle models, SIAM Multiscale Mod. Sim. 12, 3

\bibitem{KT17}
A. Klar, S. Tiwari,  A multi-scale  particle   method for mean field equations: the general case, preprint, arxiv http://arxiv.org/abs/1705.03324, 2017. 

\bibitem{kloeden1992numerical}
P. E. Kloeden and E. Platen.
\newblock {\em Numerical Solution of Stochastic Differential Equations}.
\newblock Springer-Verlag Berlin, 1992.

\bibitem{Koe14}
G. Koester, F. Treml, M. Seitz, and W. Klein, Validation of crowd models including social groups. In Ulrich Weidmann, Uwe Kirsch, and Michael Schreckenberg, editors, Pedestrian and Evacuation Dynamics 2012, 1051–1063. Springer International Publishing, 2014.


\bibitem{K14}
A.E. Kremyzas, Social Group Behavior and Path Planning, Master Thesis, University Utrecht


\bibitem{num}
H. Ling, S.C. Wong, M. Zhang, C.H. Shu, and W.H.K. Lam,
\textit{Revisiting Hughes dynamics continuum model for pedestrian flow and the development of an efficient solution algorithm},
Transp. Res. Part B: Methodological, 43 (1) (2009), pp. 127-141. 

\bibitem{maury}
B. Maury, A. Roudneff-Chupin and F. Santambrogio,
\textit{A macroscopic crowd motion model of the gradient-flow type},
Math. Models Methods Appl. Sci. 20 (2010) 1787-1921.

\bibitem{sph}
J.J. Monaghan,
\textit{Smoothed particle hydrodynamics},
Institute of physics publishing, Rep. Prog. Phys. 68 (2005), 1703-1759.
 
\bibitem{group2} 
M. Moussaid, N. Perozo, S. Garnier, D. Helbing, and G. Theraulaz,
\textit{Walking Behaviour of Pedestrian Social Groups and Its Impact on Crowd Dynamics},
[doi:10.1371/journal.pone.0010047], PLoS ONE, 5(4), e10047(2010).\\
Retrieved from http://dx.doi.org/10.1371\%2Fjournal.pone.0010047.


\bibitem{mous09} 
M. Moussaid, D. Helbing, S. Garnier, A. Johanson, M. Combe, and G. Theraulaz,
Experimental study of the behavioral mechanisms underlying self-organization in human
crowds, Proc. Roy. Soc. B Biol. Sci., 276, 2755–2762, 2009.


\bibitem{piccoli}
B. Piccoli and A. Tosin, 
\textit{Pedestrian flows in bounded domains with obstacles},
Contin. Mech. Thermodynam. 21 (2009) 85-107.

\bibitem{qui}
F. Qiu, X. Hu.
Modeling group structures in pedestrian crowd simulation.
Simulation Modelling Practice and Theory 18, 190–205, 2010.



\bibitem{eikonal1}
J.A. Sethian,
\textit{Fast marching methods},
SIAM Rev. 41 (1999) 199-235.


\bibitem{mean}
H. Spohn,
\textit{Large scale dynamics of interacting particles},
Texts and Monographs in Physics, Springer (1991).

\bibitem{SAD09}
H. Singh, R. Arter, L. Dodd, P. Langston, E. Lester, J. Drury,
Modelling subgroup behaviour in crowd dynamics DEM simulation,
Applied Mathematical Modelling,  33,  12,  2009,  4408-4423


\bibitem{tiwari}
S. Tiwari, and J. Kuhnert,
\textit{Finite pointset method based on the projection method for simulations of the incompressible Navier-Stokes equations},
Meshfree Methods for Partial Differential Equations, eds. M. Griebel and M.A. Schweitzer, Lecture Notes in Computational Science and Engineering, Vol. 26 (Springer-Verlag, 2003), pp. 373-387.

\bibitem{fpm}
S. Tiwari, and J. Kuhnert,
\textit{Modelling of two-phase flow with surface tension by finite pointset method(FPM)},
J. Comp. Appl. Math, 203 (2007), pp. 376-386.

\bibitem{TGD14}
M. Twarogowska, P. Goatin, R. Duvigneau,
Macroscopic modeling and simulations of room evacuation
Applied Mathematical Modelling, 38, Issue 24,  2014,  5781-5795

\bibitem{TCP06} 
A. Treuille, S. Cooper, Z. Popovic, 
Continuum crowds, in: ACM Transaction on Graphics, 
Proceedings of SCM SIGGRAPH 25 (2006) 1160--1168.

\bibitem{XZCH14}
J. Xi,  X. Zou, Z. Chen, J. Huang,
Multi-pattern of Complex Social Pedestrian Groups
Transportation Research Procedia
Volume 2, 2014, 60-68




\end{thebibliography}
\end{document}